\documentclass[journal]{IEEEtran}
\usepackage{algorithm}
\usepackage{algpseudocode}
\usepackage{amsmath}
\usepackage{amssymb}
\usepackage{csquotes}
\usepackage{graphicx}
\usepackage{subcaption}
\usepackage{url}

\usepackage{multirow}

\usepackage{xcolor}

\begin{document}

\title{Iterative training of neural networks for intra prediction}

\author{Thierry~Dumas, Franck~Galpin, and~Philippe~Bordes
\thanks{The authors are with Interdigital, 975 avenue des Champs Blancs, 35576, Cesson-S{\'e}vign{\'e}, France (e-mails: \protect\url{thierry.dumas@interdigital.com}, \protect\url{franck.galpin@interdigital.com}, and \protect\url{philippe.bordes@interdigital.com}).}}

\maketitle

% Abstract.
% !TeX root = iterative_training_of.tex

\begin{abstract} \label{section:abstract}
This paper presents an iterative training of neural networks for intra prediction in a block-based image and video codec. First, the neural networks are trained on blocks arising from the codec partitioning of images, each paired with its context. Then, iteratively, blocks are collected from the partitioning of images via the codec including the neural networks trained at the previous iteration, each paired with its context, and the neural networks are retrained on the new pairs. Thanks to this training, the neural networks can learn intra prediction functions that both stand out from those already in the initial codec and boost the codec in terms of rate-distortion. Moreover, the iterative process allows the design of training data cleansings essential for the neural network training. When the iteratively trained neural networks are put into H.265 (HM-16.15), $-4.2\%$ of mean BD-rate reduction is obtained, i.e. $-1.8\%$ above the state-of-the-art. By moving them into H.266 (VTM-5.0), the mean BD-rate reduction reaches $-1.9\%$.
\end{abstract}


% Introduction.
% !TeX root = iterative_training_of.tex

\section{Introduction} \label{section:introduction}
\IEEEPARstart{T}{he} prediction of an image region from a set of pixels around this region, referred to as \enquote{context}, applies to computational photography and image editing \cite{video_inpainting_of}, where it is known as \enquote{inpainting}, and video compression \cite{multiple_linear_regression}, where it is called \enquote{intra prediction}.

The prediction of an image block from its context via a machine learning approach is made difficult by the multimodal nature of the distribution $p \left( \mathbb{Y} \vert \mathbb{X} \right)$ of the random vector block $\mathbb{Y}$ given the random vector context $\mathbb{X}$. For instance, let us say that a neural network is designed to predict a block from its context, and its training consists in minimizing over its parameters the l$2$-norm of the difference between a training block and its neural network prediction, averaged over a training set of pairs of a block and its context sampled from $p \left( \mathbb{X}, \mathbb{Y} \right)$. The training assumes that $p \left( \mathbb{Y} \vert \mathbb{X} \right)$ is Gaussian with mean the neural network prediction. Besides, the training amounts to the maximum likelihood estimation of this mean \cite{harmonizing_maximum_likelihood}. After the training, as the prediction is the sample mean of the multimodal distribution, it looks blurry \cite{context_aware_semantic}.

This is commonly resolved by adversarial training \cite{generative_adversarial_nets}, which causes the learned model to pick a mode of $p \left( \mathbb{Y} \vert \mathbb{X} \right)$. For inpainting, this solution is suitable as a prediction needs to be sharp and consistent with the context \cite{generative_image_inpainting, globally_and_locally, pepsi_plus_plus}.

However, for intra prediction, adversarial training may be less appropriate. Indeed, a neural network trained in an adversarial way infers a likely prediction of a block but with potentially large pixelwise differences with it \cite{context_encoders_feature, high_resolution_image}.

\begin{figure}
	\centering
	\includegraphics[width=\linewidth]{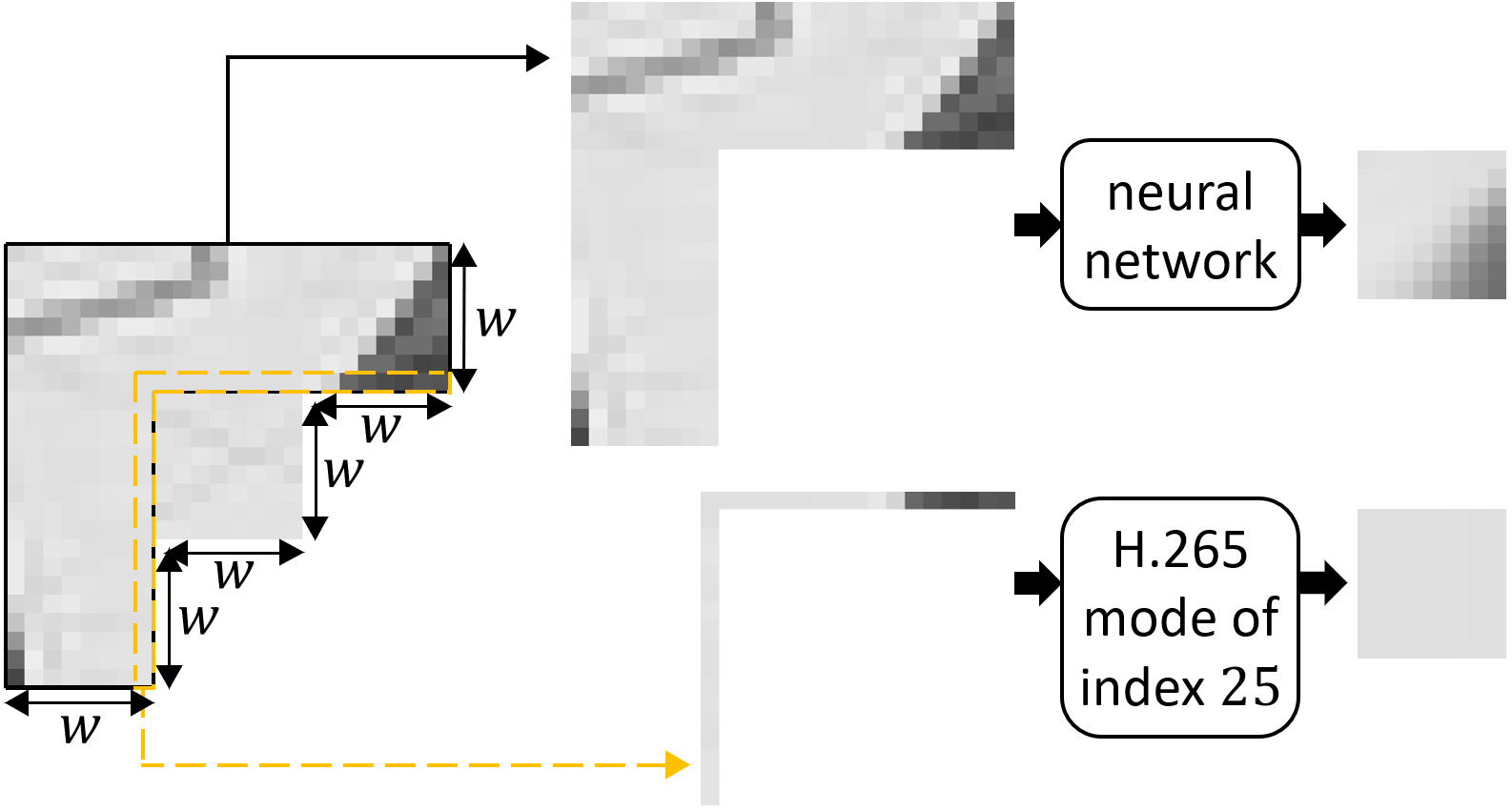}
	\caption{Prediction of a $w \times w$ luminance block returned by the H.265 (HM-16.15) partitioning of the first frame of \enquote{BQSquare} in $4$:$2$:$0$ with Quantization Parameter (QP) of $22$. At the top, a neural network trained as in \cite{fully_connected_network} predicts this block from its context of decoded pixels. At the bottom, the selected H.265 mode of index $25$ predicts this block from its reference samples, which are made of a row of $2w + 1$ decoded pixels above this block and a column of $2w$ decoded pixels on its left side. $w = 8$. The prediction PSNRs of the neural network and the H.265 mode of index $25$ are $15.82$ dB and $35.46$ dB respectively.}
	\label{figure:inference_neural_network_h265_bqsquare_qp_22}
\end{figure}

\begin{figure}
	\centering
	\includegraphics[width=\linewidth]{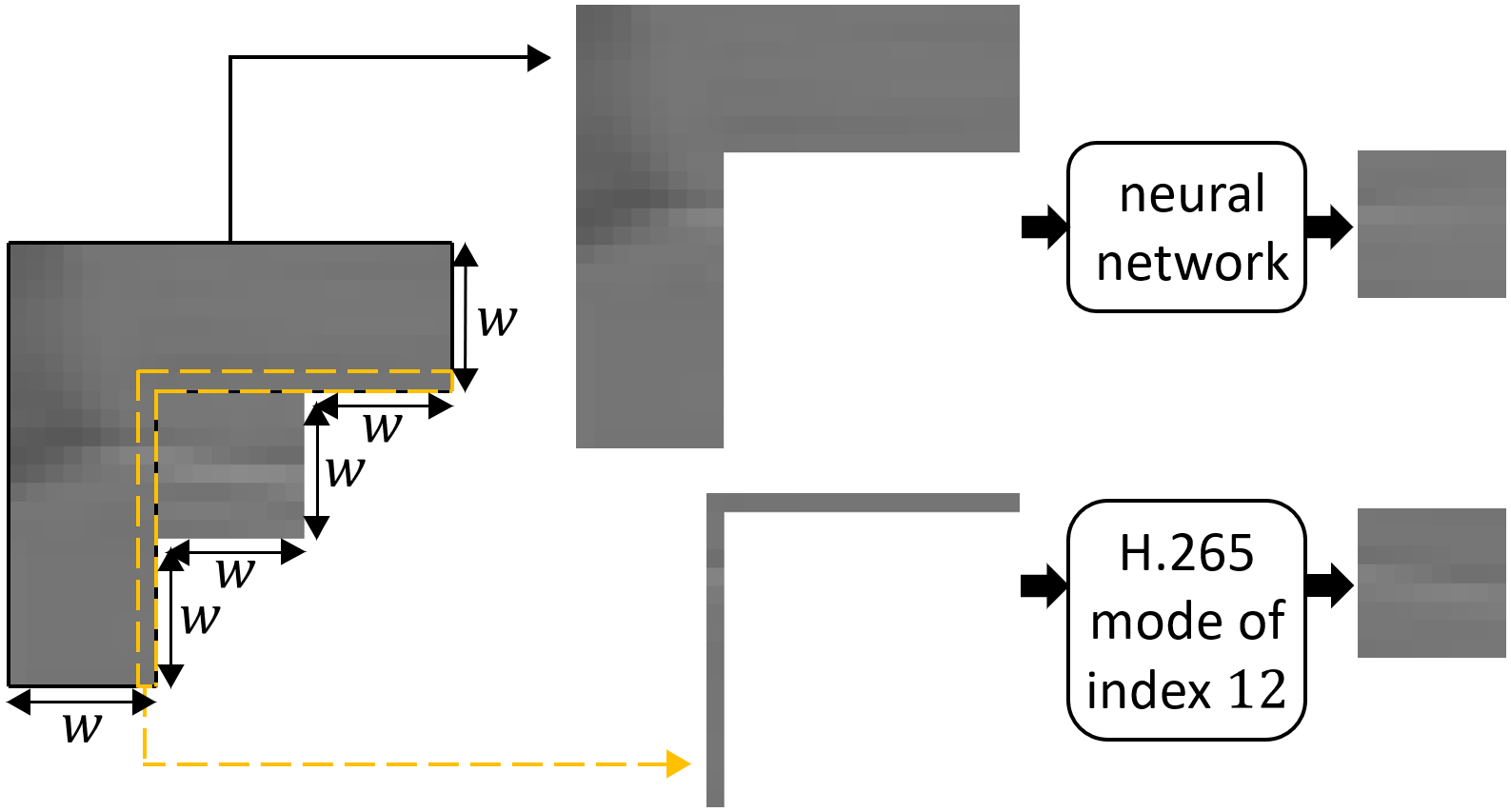}
	\caption{Prediction of a $w \times w$ luminance block returned by the H.265 (HM-16.15) partitioning of the first frame of \enquote{BasketballPass} in $4$:$2$:$0$ with $\text{QP} = 22$. At the top, a neural network trained as in \cite{fully_connected_network} predicts this block from its context of decoded pixels. At the bottom, the selected H.265 mode of index $12$ predicts this block from its reference samples. $w = 8$. The prediction PSNRs of the neural network and the H.265 mode of index $12$ are $31.90$ dB and $37.04$ dB respectively.}
	\label{figure:inference_neural_network_h265_basketballpass_qp_22}
\end{figure}

Instead, one can consider a class $\mathcal{C}$ of pairs of a block and its context defined by a known context-block relationship, then train a neural network on samples from $\mathcal{C}$, hence limiting the blurriness of the neural network predictions for $\mathcal{C}$. For example, in \cite{fully_connected_network}, $\mathcal{C}$ gathers the blocks provided by the H.265 partitioning \cite{overview_of_the} of images, each paired with its context. The context-block relationship is that a block of $\mathcal{C}$ is relatively well predicted from its reference samples by the H.265 intra prediction \cite{intra_coding_of}, see Figures~\ref{figure:inference_neural_network_h265_bqsquare_qp_22}~and~\ref{figure:inference_neural_network_h265_basketballpass_qp_22}, because the H.265 intra prediction drives the partitioning \cite{a_fast_hevc}. But, this has two drawbacks. Firstly, the neural network learns mainly the H.265 linear intra prediction. Secondly, the training set contains some pairs of a block and its context with discontinuities between the spatial distribution of pixel intensities in the block and that in its context, for which no prediction of good quality can be learned, thus hampering the training. For instance, in Figure~\ref{figure:inference_neural_network_h265_bqsquare_qp_22}, the above-mentioned discontinuities are illustrated by the diagonal dark gray border in the context not extending into the block. A neural network cannot infer this from the context \textbf{alone}, see the neural network prediction in Figure~\ref{figure:inference_neural_network_h265_bqsquare_qp_22}. Likewise, in Figure~\ref{figure:inference_neural_network_h265_basketballpass_qp_22}, the above-mentioned discontinuities are represented by the horizontal border in the context tilting downward while extending into the block, instead of extending horizontally. Again, a neural network cannot infer this from the context \textbf{alone}. In contrast, during the H.265 partitioning, these discontinuities are not problematic as the encoder considers \textbf{both} this block \textbf{and} its reference samples to select the best H.265 intra prediction mode in terms of rate-distortion.

To address the two drawbacks, an iterative training of neural networks for intra prediction in a block-based codec is proposed. At the first iteration, a set of neural networks is trained on samples from a class $\mathcal{C}$ similar to that in \cite{fully_connected_network} and put into the codec as a single additional intra prediction mode. During each successive iteration, $\mathcal{C}$ now gathers the blocks given by the image partitioning of the codec including this single additional mode, each paired with its context, and the set of neural networks is retrained on samples from $\mathcal{C}$. This way, the neural networks can learn an intra prediction progressively deviating from that in the initial codec while being beneficial in terms of rate-distortion. Moreover, from the second iteration, the pairs of a block and its context with the above-mentioned discontinuities can be detected by comparing the quality of the neural network prediction against that of the initial codec and left out of the training set.

The set of neural networks in \cite{context_adaptive_neural} is trained iteratively with H.265 as codec, then inserted into H.265 as a single additional mode, leading to $-4.2\%$ of mean BD-rate reduction. This improves on the state-of-the-art \cite{fully_connected_network, progressive_spatial_recurrent} by $-1.8\%$. When the trained neural networks are put into H.266 (VTM-5.0) as a single additional mode, the mean BD-rate reduction is $-1.9\%$.

The contributions in this paper can be summed up as:
\begin{itemize}
\item We propose an iterative training of neural networks for intra prediction.
\item A generic preprocessing of the context of a block and a generic postprocessing step are formulated to fit various block-based codecs, e.g. H.265 and H.266.
\item A signalling of the single additional neural network-based intra prediction mode in H.266 is introduced.
\end{itemize}

% Related works on intra prediction.
% !TeX root = iterative_training_of.tex

\section{Related work} \label{section:related_work}
This section presents the intra prediction in modern block-based image and video codecs. Then, it explains how some of its limitations are addressed by neural network-based methods and how our proposition falls within them.

\subsection{Intra prediction in recent block-based codecs} \label{subsection:intra_prediction_in}
Intra prediction seeks to exploit the spatial redundancies between pixels in neighboring blocks within a frame to improve coding efficiency. A block of pixels is inferred from the previously decoded neighborhood. The predicted block is subtracted from the original block, yielding a residue to be encoded. In H.265, for a given block to be predicted, one mode is selected among $35$ fixed modes according to a rate-distortion criterion. PLANAR and DC are designed to predict the blocks with gradually varying contents. PLANAR averages a horizontal propagation of the decoded neighborhood, a.k.a reference samples, and a vertical one. DC fills the predicted block with the average of the reference samples. Each of the remaining modes propagates the reference sample values into the predicted block along a specified direction \cite{intra_coding_of}. Two shortcomings of the H.265 intra prediction can be pointed out: $\left( i \right)$ as the reference samples comprise only one row and one column of decoded pixels, see Figures \ref{figure:inference_neural_network_h265_bqsquare_qp_22} and \ref{figure:inference_neural_network_h265_basketballpass_qp_22}, the intra prediction is sensitive to the quantization noise in the reference samples and cannot exploit long-range spatial dependencies, $\left( ii \right)$ the directional model of reference samples-block dependencies does not handle complex textures.

To tackle $\left( i \right)$ and $\left( ii \right)$, H.266 uses respectively Multiple Reference Lines (MRL) \cite{multiple_reference_line} and learned modes, called Matrix Intra Prediction (MIP) \cite{affine_linear_weighted}. More specifically, MRL allows to pick either the second row of decoded pixels above a luminance block to be predicted and the second column of decoded pixels on its left side or the fourth row and column instead of the first row and column. In MIP, the learned model of dependencies seems better adapted to complex textures as it mixes a directional propagation and a smooth variation in intensity. Each MIP mode predicts a $w \times h$ luminance block via a linear transformation of a downsampled version of the $w$ decoded pixels above this block and the $h$ decoded pixels on its left side. Note that, in VTM-5.0, the transformation was affine whereas, since VTM-6.0, it has been linear \cite{simplifications_of_mip}.

\subsection{Neural network-based intra prediction} \label{subsection:neural_network_based}
The neural network-based intra prediction offers alternative solutions to $\left( i \right)$ and $\left( ii \right)$. For instance, in \cite{fully_connected_network}, a neural network maps a context of several rows of decoded pixels above a block to be predicted and several columns on its left side to the predicted block, thus alleviating $\left( i \right)$. As a neural network is intended as a predictor for H.265, it is trained on blocks returned by the H.265 partitioning of images, each paired with its context. Furthermore, the authors in \cite{fully_connected_network} argue that overly complex textures should not appear in the training pairs. That is why, in a given image, each block for which the Mean Squared Error (MSE) between this block and its H.265 intra prediction is too large compared to the average prediction MSE over this image is excluded from the training set. This training data cleansing differs from ours as it does not consider the quality of the neural network prediction on a candidate training block, see Section \ref{subsection:ignoring_each_training}. Differently, in \cite{context_adaptive_neural}, the training set is filled with pairs of a block and its context extracted from images at random spatial locations. Each pair undergoes data augmentation. While the objective function to be minimized over the neural network parameters in \cite{fully_connected_network, context_adaptive_neural} is built from the MSE between a training block and its neural network prediction, the objective function in \cite{generative_adversarial_network} linearly combines the same MSE-based term with an adversarial term. The objective function in \cite{progressive_spatial_recurrent} derives from the sum of absolute transform difference between a training block and its prediction.

The originality of \cite{neural_network_based_intra, intra_picture_prediction} lies in the joint training of multiple neural networks. More precisely, blocks of each size are predicted by a different set of neural networks, and the neural network training involves a partitioner and an objective function expressed as the minimum rate-distortion cost induced by the neural network predictions over all the partitions of each training block into sub-blocks. Consequently, the set of neural networks predicting blocks of a given size specializes in a class of textures, e.g. smooth textures for a large block size. Like this training, the iterative training emphasizes some classes of textures. However, the classes differ as an iteratively trained neural network must be a generic intra predictor, see Sections \ref{subsection:generic_intra_prediction} and \ref{subsection:signalling_for_a_luminance}. Note that MIP inherits from \cite{intra_picture_prediction}.

Unlike \cite{fully_connected_network, context_adaptive_neural, progressive_spatial_recurrent, generative_adversarial_network, neural_network_based_intra, intra_picture_prediction}, and our proposition where a neural network maps the context of a block to the predicted block, and the neural network predicts this block the same way on both the encoder and decoder sides of a codec, \cite{intra_frame_prediction} splits the prediction into two parts. The neural network encoder infers from a block and its context some side information, which is sent from the encoder to the decoder. The neural network decoder computes a prediction of this block from the received side information and the context.

% Global setup of the proposed framework.
% !TeX root = iterative_training_of.tex

\section{Global setup} \label{section:global_setup}
The method developed in this paper aims at integrating a single additional neural network-based intra prediction mode into a block-based codec. Given that a neural network for intra prediction maps the L-shape context of a block to the predicted block, see Figure \ref{figure:inference_neural_network_h265_bqsquare_qp_22} and \ref{figure:inference_neural_network_h265_basketballpass_qp_22}, fully-convolutional architectures can hardly do this mapping. As full-connections are part of the architecture, the number of neural network parameters depends on the block size. Therefore, blocks of size $w \times h$ in the codec are predicted by the neural network $f_{h, w} \left( \; . \; ; \boldsymbol{\theta}_{h, w} \right)$, parametrized by $\boldsymbol{\theta}_{h, w}$, in this single additional mode. This single additional mode thus comprises $\text{card} \left( Q \right)$ neural networks, $Q$ denoting the set of possible pairs of block height and width in the codec.
\begin{figure}
	\centering
	\includegraphics[width=\linewidth]{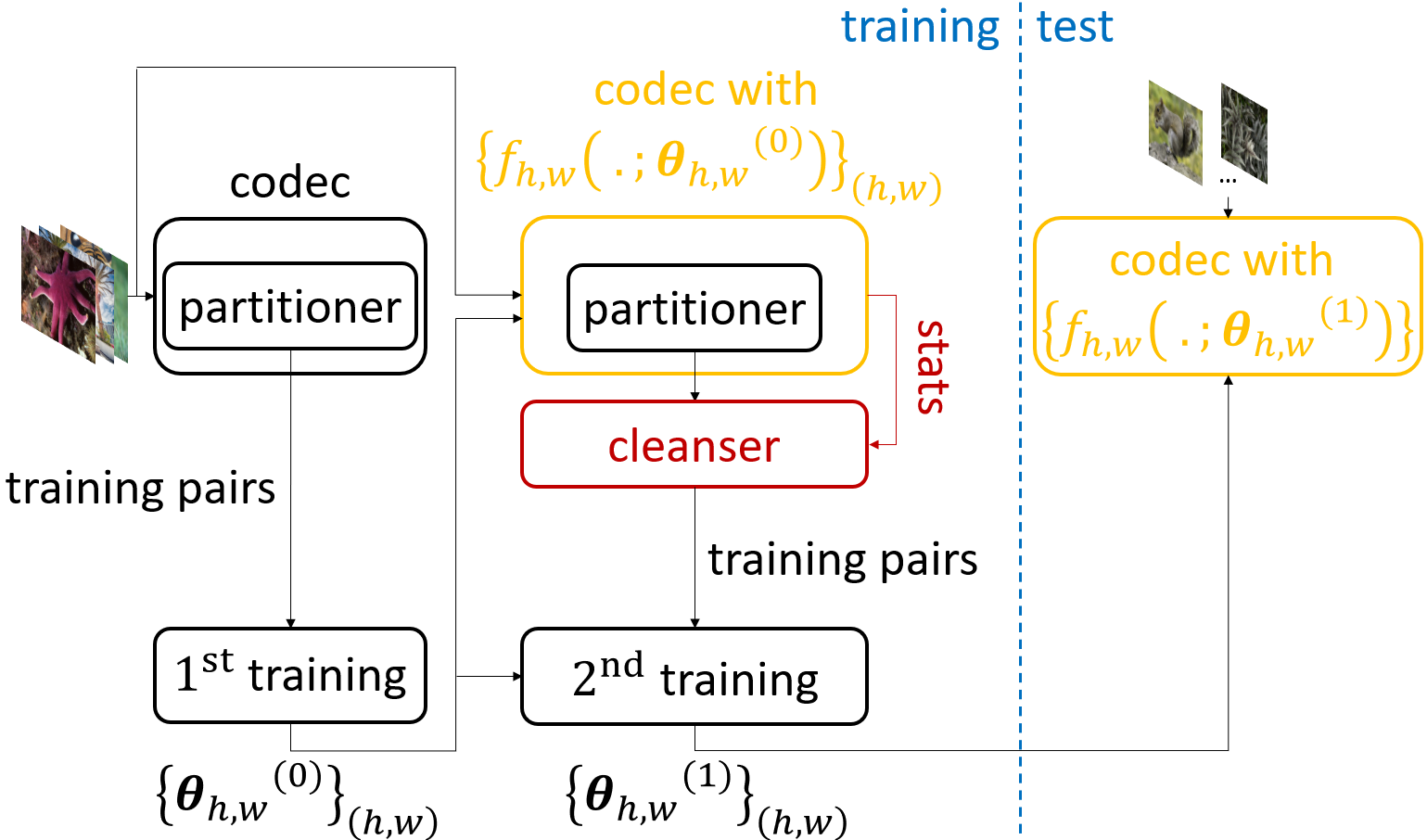}
	\caption{Evolution of the parameters of the neural networks belonging to the single additional neural network-based intra prediction mode throughout the proposed framework. Only two training iterations are displayed.}
	\label{figure:global_setup}
	
	% Final negative vertical space.
	\vspace{-3.0mm}
\end{figure}

The evolution of the parameters of these neural networks from the beginning of the iterative training to the test phase is broadly depicted in Figure \ref{figure:global_setup}. During the training phase, at each iteration, the parameters result from the neural network trainings on pairs of a block and its context extracted from the codec partitioning of images. Between the first and second iterations, the single additional mode with the parameters from the first iteration is put into the codec. Between two subsequent iterations, the single additional mode updates its parameters with the parameters from the previous iteration. During the test phase, the codec with the single additional mode using the parameters from the last training iteration encodes videos. In Figure \ref{figure:global_setup}, the neural network architectures never change.

In details, Section \ref{section:iterative_training_of} successively justifies the iterative aspect of the training, describes the creation of the training pairs from the codec partitioning of images, and explains the motivations behind the training data cleansing popping up from the second training iteration, see the red components of Figure \ref{figure:global_setup}. Then, Section \ref{section:signalling_of_the} introduces the signalling of the single additional mode in the codec, which does not vary from the second training iteration to the test phase.

% Proposed iterative training of neural networks for intra prediction.
% !TeX root = iterative_training_of.tex

\section{Iterative training of neural networks} \label{section:iterative_training_of}
This section justifies the iterative training. Some details are specified when either H.265 or H.266 is chosen as codec because they stand out as two of the most advanced video compression standards \cite{analysis_of_emerging}.

\subsection{Modification of the training data over iterations} \label{subsection:modification_of_the}
The first thrust of our approach is to avoid the case where a learned model gives blurry predictions because it was trained on an unrestricted variety of pairs of a block and its context in which many predictions of a block are likely given its context, cf. the multimodality discussed in Section \ref{section:introduction}. That is why, first, a set $\Gamma$ of $8$-bit Y$\text{C}_{\text{b}}\text{C}_{\text{r}}$ images is encoded via the codec to yield the training sets $\left\{ S_{h, w} \right\}_{\left( h, w \right) \in R}$, where $S_{h, w}$ contains pairs of a luminance block of size $w \times h$ provided by the partitioning of an image of $\Gamma$ and its context. The use of $R \subseteq Q$ instead of $Q$ will be explained two paragraphs later. Then, $f_{h, w} \left( \; . \; ; \boldsymbol{\theta}_{h, w} \right)$ is trained on $S_{h, w}$, see the first loop of Algorithm~\ref{algorithm:iterative_training_for}.
\begin{algorithm}
	\caption{: iterative training.\\ \enquote{getPartition}, \enquote{getPartitionNN}, and $\mathcal{L}_{h, w}$ are defined by Algorithm~\ref{algorithm:get_partition}, Algorithm~\ref{algorithm:get_partition_nn}, and \eqref{equation:objective_l2} respectively.}
	\label{algorithm:iterative_training_for}
	Inputs: $\Gamma$ and $l \in \mathbb{N}^{*}$
	\begin{algorithmic}
		\State{$\left\{ S_{h, w} \right\}_{\left( h, w \right) \in R} = \; \text{getPartition} \left( \Gamma \right)$}
		\ForAll{$\left( h, w \right) \in R$}
			\State{$\boldsymbol{\theta}_{h, w}^{\left( 0 \right)} = \min\limits_{\boldsymbol{\theta}_{h, w}} \; \mathcal{L}_{h, w} \left( S_{h, w}; \boldsymbol{\theta}_{h, w} \right)$ where $\boldsymbol{\theta}_{h, w}$ is}
			\State{randomly initialized.}
		\EndFor
		\ForAll{$i \in \left[ \vert 1, l - 1 \vert \right]$}
			\State{$\left\{ S_{h, w} \right\}_{\left( h, w \right) \in R} = \text{getPartitionNN} \left( \Gamma; \left\{ \boldsymbol{\theta}_{h, w}^{\left( i - 1 \right)} \right\}_{\left( h, w \right) \in R} \right)$}
			\ForAll{$\left( h, w \right) \in R$}
				\State{$\boldsymbol{\theta}_{h, w}^{\left( i \right)} = \min\limits_{\boldsymbol{\theta}_{h, w}} \; \mathcal{L}_{h, w} \left( S_{h, w}; \boldsymbol{\theta}_{h, w} \right)$ where $\boldsymbol{\theta}_{h, w} =$}
				\State{$\boldsymbol{\theta}_{h, w}^{\left( i - 1 \right)}$ at initialization.}
		\EndFor
		\EndFor
	\end{algorithmic}
	Output: $\left\{ \boldsymbol{\theta}_{h, w}^{\left( l - 1 \right)} \right\}_{\left( h, w \right) \in R}$
\end{algorithm}
\begin{algorithm}
	\caption{: getPartition.\\ \enquote{extractPair} is detailed in Figure~\ref{figure:extract_pair}. \enquote{codec} denotes the codec of interest.}
	\label{algorithm:get_partition}
	Input: $\Gamma$
	\begin{algorithmic}
		\ForAll{$\left( h, w \right) \in R$}
			\State{$S_{h, w} = \left\{ \right\}$}
		\EndFor
		\ForAll{$\mathbf{I} \in \Gamma$}
			\State{$\text{QP} \sim \mathcal{U} \left\{ 22, 27, 32, 37, 42 \right\}$}
			\State{$\hat{\mathbf{I}}, B = \text{codec} \left( \mathbf{I}, \text{QP} \right)$}
			\State{$\text{shuffle} \left( B \right)$}
			\ForAll{$\left( h, w \right) \in R$}
			    \State{$i_{h, w} = 0$}
			\EndFor
			\ForAll{$\left( x, y, h, w, n_{0}, n_{1} \right) \in B$}
			    \If{$i_{h, w} \geq q$}
					\State{\textbf{continue}}
				\EndIf
				\State{$\mathbf{X}_{c}, \mathbf{Y}_{c} = \text{extractPair} \left( \mathbf{I}, \hat{\mathbf{I}}, x, y, h, w, n_{0}, n_{1} \right)$}
				\State{$S_{h, w}.\text{append} \left( \left( \mathbf{X}_{c}, \mathbf{Y}_{c} \right) \right)$}
				\State{$i_{h, w} \mathrel{+}= 1$}
			\EndFor
		\EndFor
	\end{algorithmic}
	Output: $\left\{ S_{h, w} \right\}_{\left( h, w \right) \in R}$
\end{algorithm}
\begin{algorithm}
	\caption{: getPartitionNN\\ \enquote{extractPair} is explained in Figure~\ref{figure:extract_pair}. \enquote{codecNN} denotes the codec with the single additional neural network-based intra prediction mode.}
	\label{algorithm:get_partition_nn}
	Inputs: $\Gamma$ and $\left\{ \boldsymbol{\theta}_{h, w} \right\}_{\left( h, w \right) \in R}$
	\begin{algorithmic}
		\ForAll{$\left( h, w \right) \in R$}
			\State{$S_{h, w} = \left\{ \right\}$}
		\EndFor
		\ForAll{$\mathbf{I} \in \Gamma$}
			\State{$\text{QP} \sim \mathcal{U} \left\{ 22, 27, 32, 37, 42 \right\}$}
			\State{$\hat{\mathbf{I}}, B = \text{codecNN} \left( \mathbf{I}, \text{QP}; \left\{ \boldsymbol{\theta}_{h, w} \right\}_{\left( h, w \right) \in R} \right)$}
			\State{$\text{shuffle} \left( B \right)$}
			\ForAll{$\left( h, w \right) \in R$}
			    \State{$i_{h, w} = 0$}
			\EndFor
			\ForAll{$\left( x, y, h, w, n_{0}, n_{1}, s, d_{nn}, d_{c}, \text{isSplitTBs} \right) \in B$}
			    \If{$i_{h, w} \geq q$}
			        \State{\textbf{continue}}
			    \EndIf
				\If{$\text{isSplitTBs}$ \textbf{or} $\text{max} \left( h, w \right) \leq 4$}
					\State{$\text{isAdded} = s == s_{nn}$}
				\Else
					\State{$\text{isAdded} = d_{nn} \leq \gamma d_{c}$}
				\EndIf
				\If{$\text{isAdded}$}
					\State{$\mathbf{X}_{c}, \mathbf{Y}_{c} = \text{extractPair} \left( \mathbf{I}, \hat{\mathbf{I}}, x, y, h, w, n_{0}, n_{1} \right)$}
					\State{$S_{h, w}.\text{append} \left( \left( \mathbf{X}_{c}, \mathbf{Y}_{c} \right) \right)$}
					\State{$i_{h, w} \mathrel{+}= 1$}
				\EndIf
			\EndFor
		\EndFor
	\end{algorithmic}
	Output: $\left\{ S_{h, w} \right\}_{\left( h, w \right) \in R}$
\end{algorithm}

At this stage of the training, the learned models tend to reproduce the codec intra prediction \cite{fully_connected_network}. This results from the combination of two factors. Firstly, the context fed into a neural network always includes the reference samples fed into the codec intra prediction \cite{fully_connected_network, context_adaptive_neural, progressive_spatial_recurrent, a_hybrid_neural, combining_intra_block}, see Figures~\ref{figure:inference_neural_network_h265_bqsquare_qp_22}~and~\ref{figure:inference_neural_network_h265_basketballpass_qp_22}, implying that this neural network can learn the codec intra prediction. Secondly, the partitioning mechanism generating the training blocks ensures that each training block is \enquote{well} predicted from its reference samples by the codec intra prediction \cite{a_fast_hevc}, meaning that the codec intra prediction is a \enquote{good} solution to the learning problem for most of the training pairs. To allow the neural networks to learn an intra prediction progressively diverging from that in the codec while being valuable in terms of rate-distortion, for $l - 1$ iterations, $\left( i \right)$ training sets are built as described above, but replacing the codec by the codec with the single additional mode, $\left( ii \right)$ the neural networks are retrained on the new training sets, see the second loop of Algorithm~\ref{algorithm:iterative_training_for}. The criterion for selecting the number of training iterations $l$ will be given in Section \ref{subsubsection:benefits_of_the}.

A neural network may be needed to predict blocks of a given size in the codec but cannot be trained via Algorithm~\ref{algorithm:iterative_training_for}. This occurs when this block size exists in the intermediate steps of the partitioning, not at the output of the partitioning. For instance, a luminance Prediction Block (PB) in an intermediate step of the H.265 partitioning can be of size either $4 \times 4$, $8 \times 8$, $16 \times 16$, $32 \times 32$ or $64 \times 64$. Therefore, $Q = \left\{ \left( 4, 4 \right), \left( 8, 8 \right), \left( 16, 16 \right), \left( 32, 32 \right), \left( 64, 64 \right) \right\}$. However, the H.265 partitioning returns luminance Transform Blocks (TBs) of sizes $4 \times 4$, $8 \times 8$, $16 \times 16$, and $32 \times 32$ exclusively. Indeed, if a $64 \times 64$ luminance PB is found at a leaf of the coding tree, the split of this PB into $4$ $32 \times 32$ TBs is forced, see the second paragraph of Section 3.2.4.1 in \cite{high_efficiency_video_coding_hevc_algorithms}. Thus, $R = \left\{ \left( 4, 4 \right), \left( 8, 8 \right), \left( 16, 16 \right), \left( 32, 32 \right) \right\}$. In this case, each neural network predicting blocks of size $w \times h$, $\left( h, w \right) \in Q - R$, is trained on blocks extracted from images at random spatial locations, each paired with its context \cite{context_adaptive_neural}, see Section \ref{subsection:neural_network_based}.

\subsection{Generic intra prediction with respect to block textures} \label{subsection:generic_intra_prediction}
The single additional neural network-based intra prediction mode should predict any texture found in the blocks of the final image partitioning. This implies that the training sets must span the variety of textures in these blocks. But, large blocks from the partitioning are usually smooth whereas small blocks exhibit unsmooth textures \cite{hevc_the_new, homogeneity_based_fast}, especially at small QPs. To maintain some texture diversity in each training set, the images in $\Gamma$ should not all be encoded using a small QP. On the other hand, if the images in $\Gamma$ are all encoded using a large QP, the neural networks learn only the intra prediction from a context with large quantization noise. Therefore, the QP for encoding each image in $\Gamma$ is uniformly drawn from a set of both small and large QPs, see Algorithms~\ref{algorithm:get_partition}~and~\ref{algorithm:get_partition_nn}.

\subsection{Intra prediction from a context with missing information} \label{subsection:intra_prediction_from}
\begin{figure}
	\centering
	\includegraphics[width=0.32\linewidth]{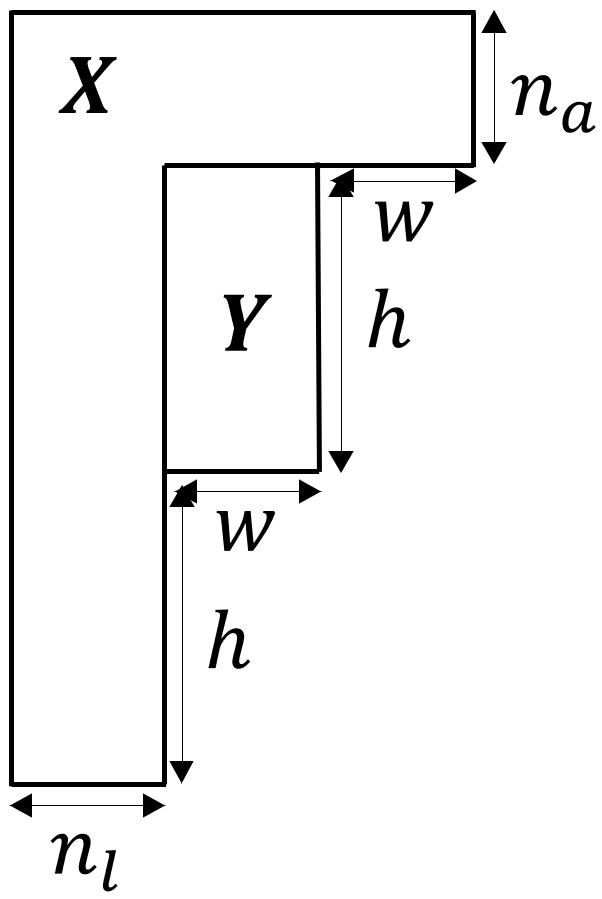}
	\caption{Positions of a $w \times h$ block $\mathbf{Y}$ and its context $\mathbf{X}$.}
	\label{figure:context_shape}
\end{figure}

In a block-based codec, the shape of the context of a block is constrained. For instance, in H.265 and H.266, as the blocks are processed in raster-scan order combined with Z-scan order, c.f. Sections 3.2.2 to 3.2.4 of \cite{high_efficiency_video_coding_hevc_algorithms} and \cite{multi_type_tree}, the context can only include decoded pixels located above and on the left side of the block. To comply with the constraints, the context $\mathbf{X}$ takes from now on the shape in Figure~\ref{figure:context_shape}. It comprises $n_{l}$ columns of $2h$ decoded pixels on the left side of the block $\mathbf{Y}$ and $n_{a}$ rows of $n_{l} + 2w$ decoded pixels above the block. As long-range spatial dependencies between decoded pixels above and on the left side of the block are needed for prediction only in the case of a large block to be predicted, a rule for defining $n_{l}$ and $n_{a}$ consists of making the ratio $\delta$ between the size of the context and that of the block constant \cite{context_adaptive_neural}. To keep a constant term in this ratio whatever $h$ and $w$ while limiting the context size, $n_{a} = n_{l} = \text{min} \left( h, w \right)$. Then,
\begin{align*}
	\delta = \text{min} \left( h, w \right) \left( \frac{\text{min} \left( h, w \right)}{hw} + \frac{2}{h} + \frac{2}{w} \right).
\end{align*}
Note that, in the case of H.265, $h = w$. As $n_{a} = n_{l} = w$, the context shape in Figure \ref{figure:context_shape} becomes that in \cite{context_adaptive_neural}.

More importantly, the block-based processing also causes information to be missing in the context of a block. Indeed, the $n_{0} \in \left[ \vert 0, h \vert \right]$ bottommost rows and/or the $n_{1} \in \left[ \vert 0 , w \vert \right]$ rightmost columns of pixels in the context may not be decoded yet, depending on the position of this block in its macro block, called Coding Tree Block (CTB) in H.265 and H.266 \cite{intra_coding_of}.
\begin{figure*}
	\centering
	\includegraphics[width=\linewidth]{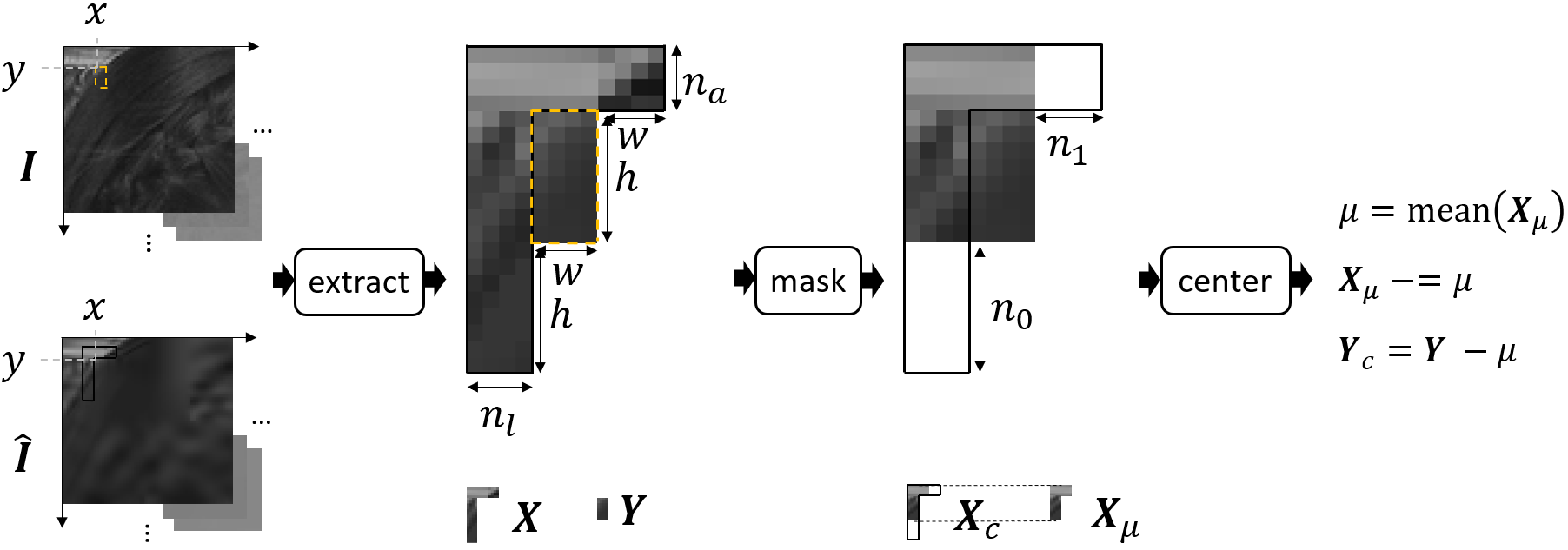}
	\caption{Creation via \enquote{extractPair} of a pair of a $w \times h$ preprocessed luminance block $\mathbf{Y}_{c}$ and its preprocessed context $\mathbf{X}_{c}$. \enquote{extractPair} gathers \enquote{extract}, \enquote{mask}, and \enquote{center}. $\mathbf{Y}$ is extracted from the luminance channel of the first frame $\mathbf{I}$ of \enquote{BlowingBubbles} in $4$:$2$:$0$ while $\mathbf{X}$ is extracted from the luminance channel of its reconstruction $\hat{\mathbf{I}}$ via H.266 (VTM-5.0) with QP = $37$. The coordinates of the pixel at the top-left of $\mathbf{Y}$ in $\mathbf{I}$ are $x = 12$ and $y = 8$. $h = 8$, $w = 4$, $n_{0} = 8$, and $n_{1} = 4$.}
	\label{figure:extract_pair}
	
	% Final negative vertical space.
	\vspace{-3.0mm}
\end{figure*}

To overcome this, the authors in \cite{progressive_spatial_recurrent} first design several contexts per block size, each context containing available decoded pixels exclusively. Then, one neural network is trained per context. During the test phase, depending on the availability of the decoded pixels around a block of a given size, the context is chosen, and its associated neural network is used for prediction. But, this increases the number of neural networks in the codec, i.e. the memory cost of their parameters.

Differently, in \cite{context_adaptive_neural}, a single context is created per block size, and the missing decoded pixels are masked with the mean pixel intensity over training luminance images. Unfortunately, the mask value belongs to the range of the available decoded pixel values, leading to ambiguities between the masked portions of the context and its unmasked portions for the neural network fed with the context.

Instead, to take the mask value out of this range, the missing decoded pixels in $\mathbf{X}$ are first covered by a mask of value $255$, see the step called \enquote{mask} in Figure~\ref{figure:extract_pair}. Then, the mean $\mu$ of the available decoded pixels is subtracted from them, yielding the preprocessed context $\mathbf{X}_{c}$. Moreover, $\mu$ is subtracted from the block $\mathbf{Y}$, giving rise to the preprocessed block $\mathbf{Y}_{c}$, see the step called \enquote{center} in Figure~\ref{figure:extract_pair}.

Now that a training pair $\left( \mathbf{X}_{c}, \mathbf{Y}_{c} \right)$ is defined, our iterative training can be further specified. In Algorithm~\ref{algorithm:get_partition}, the encoding of an image $\mathbf{I}$ of $\Gamma$ via the codec returns a reconstruction $\hat{\mathbf{I}}$ of $\mathbf{I}$ and a set $B$ of characteristics of each luminance block from the image partitioning. Then, for some of these luminance blocks, the block characteristics enable to create a training pair of a preprocessed block and its preprocessed context, see Figure~\ref{figure:extract_pair}. The same description goes for Algorithm~\ref{algorithm:get_partition_nn}, except that the codec is replaced by the codec with the single additional mode.

\subsection{Balancing the contributions of images to the training data} \label{subsection:balancing_the_contributions}
\begin{figure}
	\centering
	\includegraphics[width=0.90\linewidth]{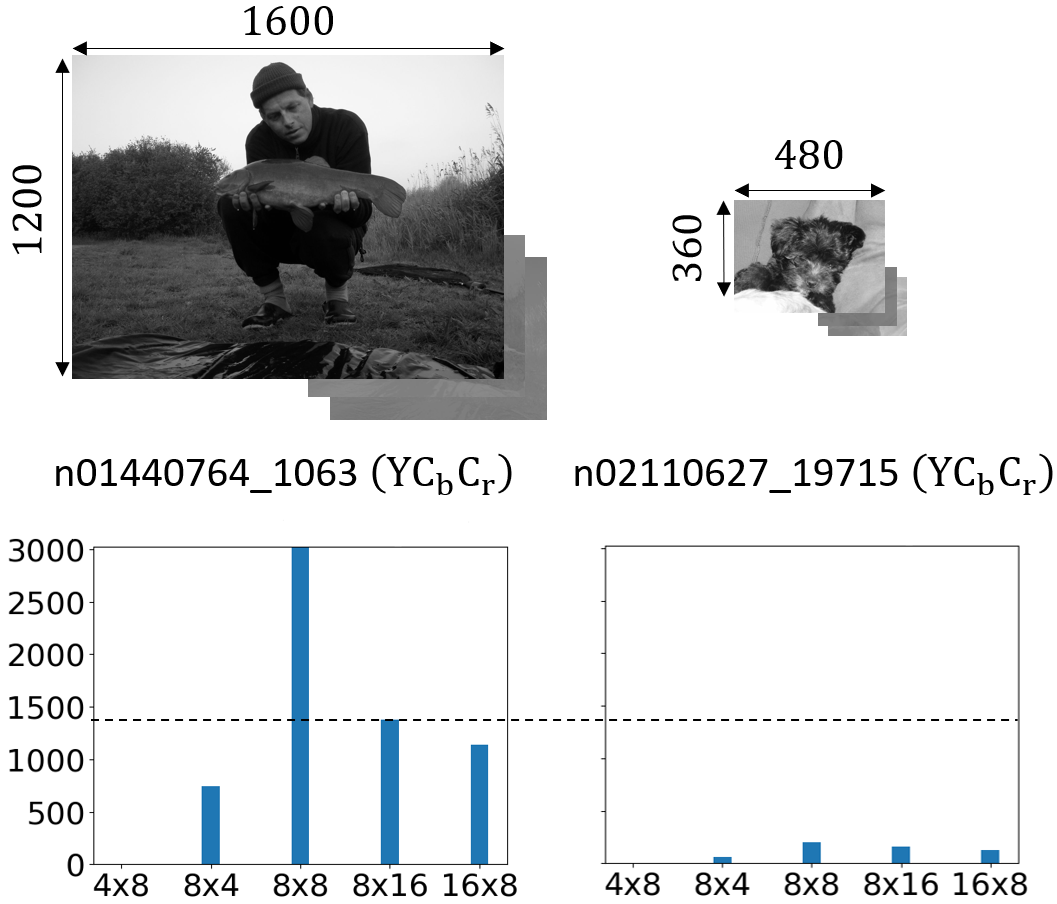}
	\caption{Number of luminance blocks of sizes $4 \times 8$, $8 \times 4$, $8 \times 8$, $8 \times 16$, and $16 \times 8$ returned by the partitionings of \enquote{n01440764\_1063} and \enquote{n02110627\_19715} in the ILSVRC2012 training RGB images \cite{imagenet_a_large}, converted into Y$\text{C}_{\text{b}}\text{C}_{\text{r}}$ and $4$:$2$:$0$, via H.266 (VTM-5.0) with $\text{QP} = 32$.}
	\label{figure:comparison_partitioning_counts}
	
	% Final negative vertical space.
	\vspace{-3.0mm}
\end{figure}

If the set $\Gamma$ contains Y$\text{C}_{\text{b}}\text{C}_{\text{r}}$ images of various sizes and, for each of them, the characteristics of each luminance block from the image partitioning give rise to a different training pair, the textures of the large images are more heavily represented in the training sets $\left\{ S_{h, w} \right\}_{\left( h, w \right) \in R}$ than those of the small images. For instance, let us focus on $S_{16, 8}$. The partitioning of the $1600 \times 1200$ image in Figure~\ref{figure:comparison_partitioning_counts} returns $1379$ $8 \times 16$ luminance blocks whereas that of the $480 \times 360$ image gives $158$ $8 \times 16$ luminance blocks. This means that almost $9$ times more training pairs in $S_{16, 8}$ come from the $1600 \times 1200$ image. To avoid this, the partitioning of an image of $\Gamma$ can add at most $q \in \mathbb{N}^{*}$ pairs of a preprocessed block and its preprocessed context to each training set $S_{h, w}$, see Algorithms~\ref{algorithm:get_partition}~and~\ref{algorithm:get_partition_nn}.

Also to build training sets with diverse textures, for each large image in $\Gamma$, the training set $S_{h, w}$ should not be filled with the first $q$ $w \times h$ preprocessed blocks coming from the top-left of this image, each paired with its preprocessed context. Thus, $B$ is shuffled before picking from it block characteristics to create training pairs, see Algorithms~\ref{algorithm:get_partition}~and~\ref{algorithm:get_partition_nn}.

\subsection{Ignoring each training block viewed as unpredictable from its context alone} \label{subsection:ignoring_each_training}
For some blocks given by the partitioning of images, the prediction of the block from its context alone via a neural network can be viewed as unfeasible, see Figures~\ref{figure:inference_neural_network_h265_bqsquare_qp_22}~and~\ref{figure:inference_neural_network_h265_basketballpass_qp_22}. If a neural network is trained on these blocks, each paired with its context, it learns to provide blurry predictions. To resolve this, these pairs must be identified and removed from the training sets. But, the identification is challenging.

A solution is to first choose a reference intra predictor that selects its prediction of a block based on \textbf{both} this block \textbf{and} some neighboring decoded pixels. Then, a given block is labeled as unpredictable from its context \textbf{alone} via a pretrained neural network if this neural network infers from the context a prediction of low quality relatively to the prediction of the reference intra predictor. The H.265 intra prediction and that in H.266 appear to be suitable reference intra predictors as the intra prediction mode for predicting a given block is selected on the encoder side by considering both this block and its reference samples. Based on this, the following basic criterion is derived. For a $w \times h$ luminance TB provided by the partitioning of an image of $\Gamma$, if the mean-squared error $d_{nn} \in \mathbb{R}_{+}^{*}$ between the luminance TB and its neural network prediction is strictly larger than $\gamma d_{c}$, $d_{c} \in \mathbb{R}_{+}^{*}$ denoting the third lowest mean-squared error between the luminance TB and the mode prediction over all the regular codec modes, this TB is not added to the training set $S_{h, w}$. By searching for the value of $\gamma \in \left\{ 0.5, 1.05, 1.55 \right\}$ yielding the best rate-distortion performance in Section \ref{subsubsection:cumulated_benefits}, $\gamma = 1.05$ was obtained.

However, the basic criterion has two downsides. Firstly, for a given luminance TB returned by the partitioning of an image, the computation of $d_{nn}$ and $d_{c}$ inside the codec increases significantly the encoding time if this TB arises from the split of its parent PB into multiple TBs. Indeed, as the intra prediction mode is defined at the PB level, the (either H.265 or H.266) encoder runs the prediction of \textbf{each} intra prediction mode on a given luminance PB to compare them\footnote{For H.265, see \enquote{TEncSearch::estIntraPredLumaQT} at \url{https://hevc.hhi.fraunhofer.de/trac/hevc/browser/trunk/source/Lib/TLibEncoder/TEncSearch.cpp}. For H.266, see \enquote{IntraSearch::estIntraPredLumaQT} at \url{https://vcgit.hhi.fraunhofer.de/jvet/VVCSoftware_VTM/blob/master/source/Lib/EncoderLib/IntraSearch.cpp}}. This implies that, if this PB is not split into multiple TBs, there is no need to run additional predictions for calculating $d_{nn}$ and $d_{c}$ for the TB equivalent to its parent PB. In contrast, when a luminance PB is split into multiple TBs, the encoder does not run the prediction of \textbf{each} mode on a child TB\footnote{For H.265, see \enquote{TEncSearch::TEncSearch::xRecurIntraCodingLumaQT} at \url{https://hevc.hhi.fraunhofer.de/trac/hevc/browser/trunk/source/Lib/TLibEncoder/TEncSearch.cpp}. For H.266, see \enquote{IntraSearch::IntraSearch::xRecurIntraCodingLumaQT} at \url{https://vcgit.hhi.fraunhofer.de/jvet/VVCSoftware_VTM/blob/master/source/Lib/EncoderLib/IntraSearch.cpp}}. In this case, the computation of $d_{nn}$ and $d_{c}$ for a child TB requires additional predictions.

To remedy to the first downside, if the luminance TB results from the split of its parent PB into multiple TBs, the basic criterion is replaced by another criterion that involves neither $d_{nn}$ nor $d_{c}$. More precisely, for a $w \times h$ luminance TB given by the partitioning of an image of $\Gamma$, if the luminance TB is equivalent to its parent PB, i.e. its flag \enquote{isSplitTBs} is false, the basic criterion applies. Otherwise, the alternative criterion is that if the index $s$ of the intra prediction mode of the luminance TB is equal to the index $s_{nn}$ of the single additional mode, this TB is added to the training set $S_{h, w}$.

As a second shortcoming, the basic criterion involves prediction PSNRs exclusively, which incurs an adverse effect for very small luminance TBs. Indeed, when several intra prediction modes yield predictions of a luminance PB with comparable qualities, the selection of the mode for predicting this PB and its child TB(s) depends on the signalling cost of each mode. This dependency increases as this PB and its child TB(s) get smaller because the first cost involved in the selection linearly combines a distortion term and a signalling term, the former growing with the block size, unlike the latter \cite{an_adaptive_lagrange, learned_fast_hevc, modified_encoder_decision}. As the basic criterion does not consider these signalling costs, there exist discrepancies between the small blocks in the training sets and those in the codec for which the single additional mode is likely to be selected.

\begin{figure}
	\centering
	\includegraphics[width=0.99\linewidth]{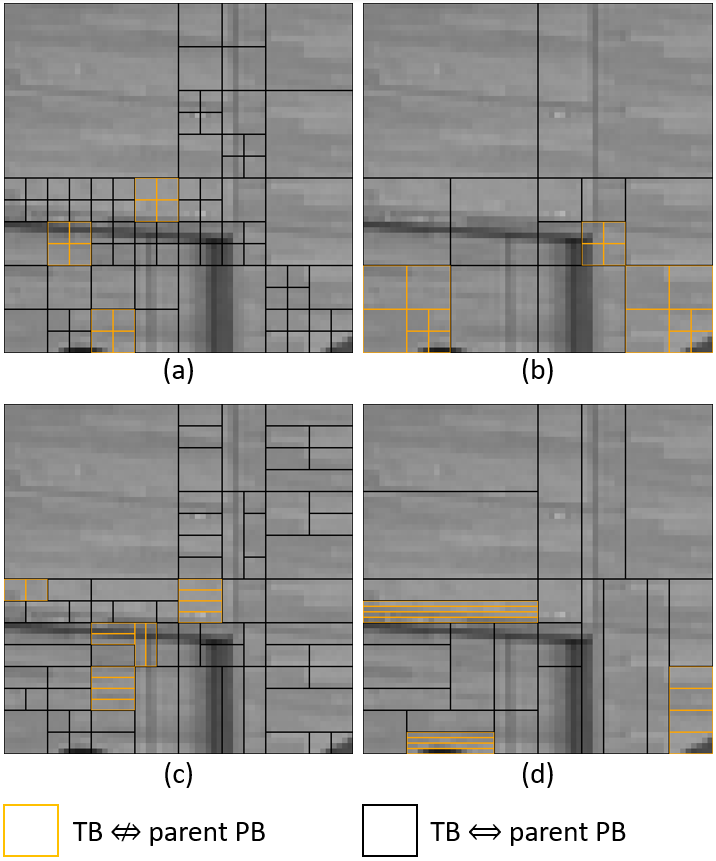}
	\caption{Partitioning of the first $64 \times 64$ block of the luminance channel of the first frame of \enquote{BasketballPass} in $4$:$2$:$0$ via H.265 (HM-16.15) with (a) $\text{QP} = 22$ and (b) $\text{QP} = 37$ and H.266 (VTM-5.0) with (c) $\text{QP} = 22$ and (d) $\text{QP} = 37$.}
	\label{figure:split_into_tbs}
	
	% Final negative vertical space.
	\vspace{-2.0mm}
\end{figure}

Unfortunately, the solution to the first downside does not tackle the second one. Indeed, although the alternative criterion takes into account the signalling cost of each intra prediction mode, the condition on \enquote{isSplitTBs} does not guarantee that the alternative criterion systematically applies to small luminance TBs. For instance, the pieces of the partitionings of the luminance channel displayed in Figure~\ref{figure:split_into_tbs} via H.265 and H.266 yield many relatively small TBs that are equivalent to their parent PB, i.e. \enquote{isSplitTBs} is false. To apply the alternative criterion to all relatively small luminance TBs, the condition for picking it over the basic criterion becomes \enquote{isSplitTBs or $\max \left( h, w \right) \leq 4$}, see Algorithm~\ref{algorithm:get_partition_nn}.

% Analysis of the iterative training.
% !TeX root = iterative_training_of.tex

\section{Analysis of the iterative training} \label{section:analysis_of_the}
Now that the proposed iterative training is fully detailed, the benefits of its two key features, namely its iterative aspect in Section \ref{subsection:modification_of_the} and its training data cleansing in Section \ref{subsection:ignoring_each_training}, must be assessed separately in terms of rate-distortion, see Section \ref{subsection:benefits_of_the}. Then, the impact of the iterative training on the behavior of the single additional neural network-based intra prediction mode in the codec is studied, see Section \ref{subsection:behavior_of_the}.

\subsection{Benefits of the two key features in terms of rate-distortion} \label{subsection:benefits_of_the}
Given that the iterative training emerges as a solution to the blurriness of the trained neural network predictions, see Section~\ref{section:introduction}, it must be evaluated by using an objective function that does not reduce this blurriness while being very efficient in intra prediction. Thus, the objective function $\mathcal{L}_{h, w}$ to be minimized over the parameters $\boldsymbol{\theta}_{h, w}$ of $f_{h, w} \left( \; . \; ; \boldsymbol{\theta}_{h, w} \right)$ is built on the l$2$-norm of the difference between a $w \times h$ block and its neural network prediction. Moreover, regularization via l$2$-norm of the neural network weights $\mathbf{W}_{h, w}$ applies \cite{practical_recommendations_for}.
\begin{equation} \label{equation:objective_l2}
\begin{split}
	\mathcal{L} \left( S_{h, w}; \boldsymbol{\theta}_{h, w} \right) = &\frac{1}{\text{card} \left( S_{h, w} \right)} \sum\limits_{\left( \mathbf{X}_{c}, \mathbf{Y}_{c} \right) \in S_{h, w}} \left\Vert \mathbf{Y}_{c} - \hat{\mathbf{Y}}_{c} \right\Vert_{2}\\
	&+ \lambda \Vert \mathbf{W}_{h, w} \Vert_{2}^{2}
\end{split}
\end{equation}
where $\hat{\mathbf{Y}}_{c} = f_{h, w} \left( \mathbf{X}_{c}; \boldsymbol{\theta}_{h, w} \right)$ and $\lambda = 0.0005$. $\Gamma$ gathers the ILSVRC2012 \cite{imagenet_a_large} and DIV2K \cite{ntire_2017_challenge} training RGB images converted into Y$\text{C}_{\text{b}}\text{C}_{\text{r}}$. As $\Gamma$ contains nearly $1.3 \times 10^{6}$ images and it is observed experimentally that around $20 \times 10^{6}$ pairs per training set prevents overfitting, the maximum number $q$ of pairs the partitioning of an image brings to each training set, see Algorithms \ref{algorithm:get_partition} and \ref{algorithm:get_partition_nn}, is fixed to $20$.

As this paper does not address the neural network architectures, the experiments reuse the neural networks architectures in \cite{context_adaptive_neural}. More precisely, the neural network predicting $w \times h$ blocks with $\min \left( h, w \right) \leq 8$ is made of four fully-connected layers. Each layer excluding the last one has $1200$ output neurons. The last layer has $hw$ output neurons. The neural network predicting $w \times h$ blocks with $\min \left( h, w \right) > 8$ takes the convolutional architecture in \cite{context_adaptive_neural} with a stack of $4$ convolutional layers for each of the two rectangular portions of the context in Figure \ref{figure:context_shape}, a channelwise fully-connected layer, and a stack of $4$ transpose convolutional layers. The number of output feature maps in the $4$ convolutional layers are respectively $64$, $64$, $128$, and $128$. The number of output feature maps in the $4$ transpose convolutional layers are respectively $128$, $64$, $64$, and $1$. In all architectures, each layer apart from the last one has LeakyReLU with $0.1$ as slope. Please refer to Table XIII (c) and (d) in \cite{context_adaptive_neural} to know the stride and the kernel size in each layer of the convolutional architecture. Given the above description and Figure \ref{figure:context_shape}, the four neural networks predicting blocks of sizes $4 \times 4$, $8 \times 8$, $16 \times 16$, and $32 \times 32$ contain respectively $2998816$, $3344464$, $1339072$, and $3802816$ parameters. Moreover, the distributions for initializing the neural network parameters and the training hyperparameters in \cite{context_adaptive_neural} are re-used. The training involves Tensorflow 1.9.0 \cite{tensorflow_a_system} and a GPU NVIDIA Tesla P100.

Up to now, only the training phase of the neural networks has been covered. For the following experiments, details on the test phase must be specified. During the test phase, for a given $w \times h$ block to be predicted, the preprocessing of its context and the postprocessing of the neural network prediction derive from the context preprocessing during the training phase. A single step is added to adjust to the codec internal bitdepth $b \in \left\{ 8, 10 \right\}$. More specifically, the context $\mathbf{X}$ of this block is divided by $2^{b - 8}$ and preprocessed as described by \enquote{mask} and \enquote{center} in Figure~\ref{figure:extract_pair}, yielding the preprocessed context $\mathbf{X}_{c}^{t}$. Then, the neural network prediction is postprocessed by adding the mean $\mu$ of the available decoded pixels in $\mathbf{X}$, multiplying by $2^{b - 8}$, and clipping to $\left[ 0, 2^{b} - 1 \right]$, yielding the prediction
\begin{align*} \label{equation:preprocessing_postprocessing}
	\hat{\mathbf{Y}} = \text{min} \left( \text{max} \left( 2^{b - 8} \left( f_{h, w} \left( \mathbf{X}_{c}^{t}; \boldsymbol{\theta}_{h, w} \right) + \mu \right), 0 \right), 2^{b} - 1 \right).
\end{align*}

Section \ref{subsection:benefits_of_the} uses H.265 (HM-16.15). The signalling of the single additional mode in H.265 follows \cite{context_adaptive_neural}. H.265 with the single additional mode is denoted H.265-ITNN, ITNN standing for Iteratively Trained Neural Networks. The test set gathers luminance images only as the neural networks are trained on pairs of a luminance block and its context and, to simplify the interpretation of the results, the intra prediction of chrominance blocks is temporarily set aside. Therefore, H.265-ITNN and H.265 encode the luminance channel of the first frame of video sequences from the Common Test Conditions (CTC) \cite{common_test_conditions}. Note that the neural networks cannot specialize to the CTC video sequences as they do not belong to $\Gamma$. The rate-distortion performance is computed via the Bjontegaard metric \cite{calculation_of_average} of H.265-ITNN with respect to H.265 with $\text{QP} \in \left\{ 17, 22, 27, 32, 37, 42 \right\}$.

\subsubsection{Benefits of the iterative aspect} \label{subsubsection:benefits_of_the}
To avoid mixing up the contributions of the iterative aspect and the training data cleansing to the rate-distortion performance, the BD-rate reductions in Table \ref{table:benefit_of_the_iterative} are computed by eliminating the training data cleansing from the iterative training. This means that Algorithm~\ref{algorithm:get_partition_nn} becomes Algorithm~\ref{algorithm:get_partition}, but substituting \enquote{codec} with \enquote{codecNN}.
\setlength{\tabcolsep}{5.4pt}
\begin{table}
	\caption{BD-rate reductions in $\%$ of H.265-ITNN after each iteration of the iterative training without the training data cleansing. The anchor is H.265 (HM-16.15). Only the luminance channel of the first frame of each video sequence is considered. $\left( l, p \right)$ refers to the training using $l$ iterations and, at each iteration, multiplying by $p \in \mathbb{N}^{*}$ the number of training iterations at each stage of the learning rate scheduling.}
	\centering
	\begin{tabular}{llccccc}
		\hline
		\multicolumn{2}{l}{\multirow{2}{*}{Video sequence}} & \multicolumn{5}{c}{BD-rate reduction of H.265-ITNN}\\ \cline{3-7} & & $\left( 1, 2 \right)$ & $\left( 1, 3 \right)$ & $\left( 1, 1 \right)$ & $\left( 2, 1 \right)$ & $\left( 3, 1 \right)$\\
		\hline
		\multirow{8}{*}{A} & CampfireParty & $-3.32$ & $-3.32$ & $-3.38$ & $-3.49$ & $-3.58$\\ & DaylightRoad2 & $-3.96$ & $-3.98$ & $-3.95$ & $-4.26$ & $-4.51$\\ & Drums2 & $-4.68$ & $-4.71$ & $-4.67$ & $-5.00$ & $-5.03$\\ & Tango2 & $-6.36$ & $-6.39$ & $-6.36$ & $-6.40$ & $-6.54$ \\ & ToddlerFountain2 & $-3.30$ & $-3.40$ & $-3.36$ & $-3.48$ & $-3.55$\\ & TrafficFlow & $-4.29$ & $-4.42$ & $-4.48$ & $-4.74$ & $-4.91$\\ & PeopleOnStreet & $-5.82$ & $-5.81$ & $-5.99$ & $-6.15$ & $-6.14$\\ & Traffic & $-4.69$ & $-4.71$ & $-4.74$ & $-4.96$ & $-5.04$\\
		\hline
		\multirow{5}{*}{B} & BasketballDrive & $-7.09$ & $-7.14$ & $-7.25$ & $-8.07$ & $-8.37$\\ & BQTerrace & $-3.81$ & $-3.79$ & $-4.09$ & $-4.66$ & $-4.75$\\ & Cactus & $-4.15$ & $-4.08$ & $-4.22$ & $-4.61$ & $-4.78$\\ & Kimono & $-3.89$ & $-3.89$ & $-3.89$ & $-4.04$ & $-4.08$\\ & ParkScene & $-2.82$ & $-2.82$ & $-2.91$ & $-2.91$ & $-2.95$\\
		\hline
		\multirow{4}{*}{C} & BasketballDrill & $-4.68$ & $-4.70$ & $-4.81$ & $-5.12$ & $-5.24$\\ & BQMall & $-3.93$ & $-3.92$ & $-3.96$ & $-4.37$ & $-4.56$\\ & PartyScene & $-2.98$ & $-2.99$ & $-2.94$ & $-3.19$ & $-3.36$\\ & RaceHorses & $-3.76$ & $-3.73$ & $-3.77$ & $-4.12$ & $-4.13$\\
		\hline
		\multirow{4}{*}{D} & BasketballPass & $-4.49$ & $-4.52$ & $-4.10$ & $-4.71$ & $-4.88$\\ & BlowingBubbles & $-3.17$ & $-3.09$ & $-3.25$ & $-3.48$ & $-3.52$\\ & BQSquare & $-3.47$ & $-3.46$ & $-3.38$ & $-3.78$ & $-3.78$\\ & RaceHorses & $-4.97$ & $-4.96$ & $-4.96$ & $-5.29$ & $-5.50$\\
		\hline
		\multicolumn{2}{l}{Mean} & $-4.27$ & $-4.28$ & $-4.31$ & $-4.61$ & $-4.72$\\
		\hline 
	\end{tabular}
	\label{table:benefit_of_the_iterative}
	\vspace{-2.0mm}
\end{table}

The second training iteration yields $-0.30\%$ of additional mean BD-rate reduction with respect to the first one. From the second iteration to the third one, $-0.11\%$ of additional mean BD-rate reduction is obtained, see the last three columns of Table \ref{table:benefit_of_the_iterative}. The benefits of the iterative training come from a change in the training data, not an underfitting at the first iteration. This is proved by two experimental pieces of evidence. Firstly, when $l = 1$ and the number of training iterations at each stage of the learning rate scheduling is multiplied by $p \in \left\{ 2, 3 \right\}$, almost no change in the mean BD-rate reduction is reported with respect to the case $l = 1$ and $p = 1$, see the first three columns of Table \ref{table:benefit_of_the_iterative}. Yet, the case $l = 1$ and $p = 3$ and the case $l = 3$ and $p = 1$ amount to the same number of training iterations. Besides, we have run the case $l = 3$ and $p = 1$ with the random initialization of the neural network parameters at each iteration. This means that, in Algorithm~\ref{algorithm:iterative_training_for}, the random initialization of the first minimization applies to all subsequent minimizations. In this case, negligible variations in BD-rate reductions with respect to those in the last column of Table \ref{table:benefit_of_the_iterative} have been observed.

When a fourth training iteration was run, $-0.02\%$ of additional mean BD-rate reduction with respect to the third iteration was obtained. As the rate-distortion benefit of the fourth training iteration is viewed as small compared to those of the first three iterations, $l = 3$ in Sections \ref{subsection:behavior_of_the} and \ref{section:comparison_with_the}.

\subsubsection{Benefits of the training data cleansing}
\setlength{\tabcolsep}{8pt}
\begin{table}
	\caption{BD-rate reductions in $\%$ of H.265-ITNN when the training data cleansing is turned off and on. The anchor is H.265 (HM-16.15). Only the luminance channel of the first frame of each video sequence is considered.}
	\centering
	\begin{tabular}{llcccc}
		\hline
		\multicolumn{2}{l}{\multirow{3}{*}{Video sequence}} & \multicolumn{4}{c}{BD-rate reduction of H.265-ITNN}\\ \cline{3-6} & & \multicolumn{2}{c}{cleansing off} & \multicolumn{2}{c}{cleansing on}\\ \cline{3-6} & & $\left( 2, 1 \right)$ & $\left( 3, 1 \right)$ & $\left( 2, 1 \right)$ & $\left( 3, 1 \right)$\\
		\hline
		\multirow{8}{*}{A} & CampfireParty & $-3.49$ & $-3.58$ & $-3.62$ & $-3.67$\\ & DaylightRoad2 & $-4.26$ & $-4.51$ & $-4.66$ & $-4.63$\\ & Drums2 & $-5.00$ & $-5.03$ & $-5.22$ & $-5.33$\\ & Tango2 & $-6.40$ & $-6.54$ & $-6.52$ & $-6.54$ \\ & ToddlerFountain2 & $-3.48$ & $-3.55$ & $-3.58$ & $-3.67$\\ & TrafficFlow & $-4.74$ & $-4.91$ & $-5.21$ & $-5.40$\\ & PeopleOnStreet & $-6.15$ & $-6.14$ & $-6.40$ & $-6.47$\\ & Traffic & $-4.96$ & $-5.04$ & $-5.32$ & $-5.43$\\
		\hline
		\multirow{5}{*}{B} & BasketballDrive & $-8.07$ & $-8.37$ & $-8.83$ & $-8.85$\\ & BQTerrace & $-4.66$ & $-4.75$ & $-5.26$ & $-5.38$\\ & Cactus & $-4.61$ & $-4.78$ & $-5.10$ & $-5.10$\\ & Kimono & $-4.04$ & $-4.08$ & $-4.20$ & $-4.19$\\ & ParkScene & $-2.91$ & $-2.95$ & $-3.01$ & $-3.12$\\
		\hline
		\multirow{4}{*}{C} & BasketballDrill & $-5.12$ & $-5.24$ & $-6.48$ & $-6.73$\\ & BQMall & $-4.37$ & $-4.56$ & $-5.15$ & $-5.18$\\ & PartyScene & $-3.19$ & $-3.36$ & $-3.57$ & $-3.71$\\ & RaceHorses & $-4.12$ & $-4.13$ & $-4.57$ & $-4.55$\\
		\hline
		\multirow{4}{*}{D} & BasketballPass & $-4.71$ & $-4.88$ & $-6.31$ & $-6.35$\\ & BlowingBubbles & $-3.48$ & $-3.52$ & $-4.06$ & $-4.09$\\ & BQSquare & $-3.78$ & $-3.78$ & $-4.35$ & $-4.50$\\ & RaceHorses & $-5.29$ & $-5.50$ & $-5.57$ & $-5.89$\\
		\hline
		\multicolumn{2}{l}{Mean} & $-4.61$ & $-4.72$ & $-5.09$ & $-5.18$\\
		\hline 
	\end{tabular}
	\label{table:benefit_of_the_training}
\end{table}
Table \ref{table:benefit_of_the_training} shows that, for $l = 2$ and $p = 1$, $-0.48\%$ of additional mean BD-rate reduction is reported when the training data cleansing is turned on with respect to the case where it is turned off. For $l = 3$ and $p = 1$, the additional mean BD-rate reduction reaches $-0.46\%$. Therefore, the iterative aspect and the training data cleansing have equally large rate-distortion impacts.

\subsubsection{Cumulated benefits} \label{subsubsection:cumulated_benefits}
Overall, the iterative training yields $-0.87\%$ of additional mean BD-rate reduction compared to a single-step training, see the third column of Table \ref{table:benefit_of_the_iterative} and the last column of Table \ref{table:benefit_of_the_training}.

\subsubsection{Influence of the training QPs}
The experiments in Section \ref{subsubsection:benefits_of_the} have also been re-run by replacing the set of QPs in Algorithms \ref{algorithm:get_partition} and \ref{algorithm:get_partition_nn} with $\left\{ 17, 24 \right\}$. The mean BD-rate reduction for $l = 3$ with $\left\{ 17, 24 \right\}$ as set of QPs is smaller in absolute value than that with $\left\{ 22, 27, 32, 37, 42 \right\}$ as set of QPs by nearly $0.03$. When re-running the experiments by replacing the set of QPs in Algorithms \ref{algorithm:get_partition} and \ref{algorithm:get_partition_nn} with $\left\{ 35, 40 \right\}$, the mean BD-rate reduction for $l = 3$ with $\left\{ 35, 40 \right\}$ as set of QPs is smaller in absolute value than that with $\left\{ 22, 27, 32, 37, 42 \right\}$ as set of QPs by $0.40$. This validates the choice of the set of both small and large QPs in Algorithms \ref{algorithm:get_partition} and \ref{algorithm:get_partition_nn}.

\subsection{Behavior of the modified codec through iterations} \label{subsection:behavior_of_the}
The rate-distortion benefits identified in Section \ref{subsubsection:cumulated_benefits} lead to the question of how the iterative training affects the codec with the single additional mode.

To answer this, let us first analyze how the neural network prediction evolves over iterations. This analysis requires to predict a given block from its context via a neural network after two different training iterations. But, within H.265-ITNN, the quantization noise in the context of this block varies over iterations as the codec intra prediction changes over iterations, making the predictions incomparable. That is why the first analysis is not performed inside H.265-ITNN. Instead, the following experiment acts as a substitute. Triplets of a $w \times w$ block, its context, and its reference samples are extracted from the luminance channel of Y$\text{C}_{\text{b}}\text{C}_{\text{r}}$ images at random spatial locations. These images are obtained by converting into Y$\text{C}_{\text{b}}\text{C}_{\text{r}}$ the $24$ RGB images in the Kodak suite \cite{kodak_suite} and the $100$ RGB images in the BSDS test dataset \cite{a_database_of}. Then, for each triplet, the two predictions of the block given by the neural network $f_{w, w} \left( . \; ; \boldsymbol{\theta}_{w, w}^{\left( i \right)} \right)$ after the first and third training iterations, i.e. $i \in \left\{ 0, 2 \right\}$, are compared to the prediction of the best H.265 intra prediction mode in terms of prediction PSNR.

\begin{figure}
	\centering
	\begin{subfigure}{0.24\linewidth}
		\centering
		\includegraphics[width=\linewidth]{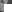}
		\caption{}
	\end{subfigure}
	\hspace{2.0mm}
	\begin{subfigure}{0.080\linewidth}
		\centering
		\includegraphics[width=\linewidth]{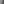}
		\caption{}
	\end{subfigure}
	\hspace{2.0mm}
	\begin{subfigure}{0.080\linewidth}
		\centering
		\includegraphics[width=\linewidth]{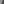}
		\caption{}
	\end{subfigure}
	\hspace{2.0mm}
	\begin{subfigure}{0.165\linewidth}
		\centering
		\vspace{5.2mm}
		\includegraphics[width=\linewidth]{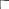}
		\caption{}
	\end{subfigure}
	\hspace{2.0mm}
	\begin{subfigure}{0.080\linewidth}
		\centering
		\includegraphics[width=\linewidth]{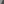}
		\caption{}
	\end{subfigure}
	\hspace{2.0mm}
	\begin{subfigure}{0.080\linewidth}
		\centering
		\includegraphics[width=\linewidth]{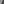}
		\caption{}
	\end{subfigure}
	\caption{Predictions of a $4 \times 4$ luminance block: (a) context, (b) prediction from the context via the neural network after the first training iteration ($\text{PSNR} = 23.74$ dB), (c) prediction via the neural network after the third iteration ($\text{PSNR} = 30.44$ dB), (d) reference samples, (e) prediction from the reference samples via the best H.265 mode of index $30$ in terms of prediction PSNR ($\text{PSNR} = 28.79$ dB), and (f) original block.}
	\label{figure:predictions_kodak_4_4_418}
\end{figure}

\begin{figure}
	\centering
	\begin{subfigure}{0.24\linewidth}
		\centering
		\includegraphics[width=\linewidth]{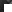}
		\caption{}
	\end{subfigure}
	\hspace{2.0mm}
	\begin{subfigure}{0.080\linewidth}
		\centering
		\includegraphics[width=\linewidth]{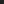}
		\caption{}
	\end{subfigure}
	\hspace{2.0mm}
	\begin{subfigure}{0.080\linewidth}
		\centering
		\includegraphics[width=\linewidth]{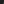}
		\caption{}
	\end{subfigure}
	\hspace{2.0mm}
	\begin{subfigure}{0.165\linewidth}
		\centering
		\vspace{5.2mm}
		\includegraphics[width=\linewidth]{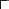}
		\caption{}
	\end{subfigure}
	\hspace{2.0mm}
	\begin{subfigure}{0.080\linewidth}
		\centering
		\includegraphics[width=\linewidth]{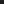}
		\caption{}
	\end{subfigure}
	\hspace{2.0mm}
	\begin{subfigure}{0.080\linewidth}
		\centering
		\includegraphics[width=\linewidth]{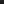}
		\caption{}
	\end{subfigure}
	\caption{Predictions of a $4 \times 4$ luminance block: (a) context, (b) prediction from the context via the neural network after the first training iteration ($\text{PSNR} = 33.60$ dB), (c) prediction via the neural network after the third iteration ($\text{PSNR} = 30.66$ dB), (d) reference samples, (e) prediction from the reference samples via the best H.265 mode of index $10$ in terms of prediction PSNR ($\text{PSNR} = 38.30$ dB), and (f) original block.}
	\label{figure:predictions_kodak_4_4_688}
\end{figure}

\begin{figure}
	\centering
	\begin{subfigure}{0.24\linewidth}
		\centering
		\includegraphics[width=\linewidth]{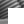}
		\caption{}
	\end{subfigure}
	\hspace{2.0mm}
	\begin{subfigure}{0.080\linewidth}
		\centering
		\includegraphics[width=\linewidth]{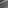}
		\caption{}
	\end{subfigure}
	\hspace{2.0mm}
	\begin{subfigure}{0.080\linewidth}
		\centering
		\includegraphics[width=\linewidth]{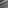}
		\caption{}
	\end{subfigure}
	\hspace{2.0mm}
	\begin{subfigure}{0.165\linewidth}
		\centering
		\vspace{5.2mm}
		\includegraphics[width=\linewidth]{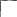}
		\caption{}
	\end{subfigure}
	\hspace{2.0mm}
	\begin{subfigure}{0.080\linewidth}
		\centering
		\includegraphics[width=\linewidth]{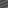}
		\caption{}
	\end{subfigure}
	\hspace{2.0mm}
	\begin{subfigure}{0.080\linewidth}
		\centering
		\includegraphics[width=\linewidth]{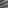}
		\caption{}
	\end{subfigure}
	\caption{Predictions of a $8 \times 8$ luminance block: (a) context, (b) prediction from the context via the neural network after the first training iteration ($\text{PSNR} = 23.71$ dB), (c) prediction via the neural network after the third iteration ($\text{PSNR} = 25.76$ dB), (d) reference samples, (e) prediction from the reference samples via the best H.265 mode of index $6$ in terms of prediction PSNR ($\text{PSNR} = 20.94$ dB), and (f) original block.}
	\label{figure:predictions_kodak_8_8_193}
\end{figure}

\begin{figure}
	\centering
	\begin{subfigure}{0.24\linewidth}
		\centering
		\includegraphics[width=\linewidth]{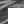}
		\caption{}
	\end{subfigure}
	\hspace{2.0mm}
	\begin{subfigure}{0.080\linewidth}
		\centering
		\includegraphics[width=\linewidth]{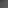}
		\caption{}
	\end{subfigure}
	\hspace{2.0mm}
	\begin{subfigure}{0.080\linewidth}
		\centering
		\includegraphics[width=\linewidth]{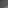}
		\caption{}
	\end{subfigure}
	\hspace{2.0mm}
	\begin{subfigure}{0.165\linewidth}
		\centering
		\vspace{5.2mm}
		\includegraphics[width=\linewidth]{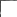}
		\caption{}
	\end{subfigure}
	\hspace{2.0mm}
	\begin{subfigure}{0.080\linewidth}
		\centering
		\includegraphics[width=\linewidth]{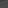}
		\caption{}
	\end{subfigure}
	\hspace{2.0mm}
	\begin{subfigure}{0.080\linewidth}
		\centering
		\includegraphics[width=\linewidth]{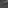}
		\caption{}
	\end{subfigure}
	\caption{Predictions of a $8 \times 8$ luminance block: (a) context, (b) prediction from the context via the neural network after the first training iteration ($\text{PSNR} = 24.45$ dB), (c) prediction via the neural network after the third iteration ($\text{PSNR} = 22.24$ dB), (d) reference samples, (e) prediction from the reference samples via the best H.265 mode of index $7$ in terms of prediction PSNR ($\text{PSNR} = 27.63$ dB), and (f) original block.}
	\label{figure:predictions_kodak_8_8_209}
\end{figure}

\begin{figure}
	\centering
	\begin{subfigure}{0.24\linewidth}
		\centering
		\includegraphics[width=\linewidth]{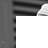}
		\caption{}
	\end{subfigure}
	\hspace{2.0mm}
	\begin{subfigure}{0.080\linewidth}
		\centering
		\includegraphics[width=\linewidth]{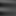}
		\caption{}
	\end{subfigure}
	\hspace{2.0mm}
	\begin{subfigure}{0.080\linewidth}
		\centering
		\includegraphics[width=\linewidth]{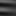}
		\caption{}
	\end{subfigure}
	\hspace{2.0mm}
	\begin{subfigure}{0.165\linewidth}
		\centering
		\vspace{5.2mm}
		\includegraphics[width=\linewidth]{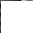}
		\caption{}
	\end{subfigure}
	\hspace{2.0mm}
	\begin{subfigure}{0.080\linewidth}
		\centering
		\includegraphics[width=\linewidth]{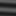}
		\caption{}
	\end{subfigure}
	\hspace{2.0mm}
	\begin{subfigure}{0.080\linewidth}
		\centering
		\includegraphics[width=\linewidth]{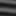}
		\caption{}
	\end{subfigure}
	\caption{Predictions of a $16 \times 16$ luminance block: (a) context, (b) prediction from the context via the neural network after the first training iteration ($\text{PSNR} = 22.85$ dB), (c) prediction via the neural network after the third iteration ($\text{PSNR} = 26.09$ dB), (d) reference samples, (e) prediction from the reference samples via the best H.265 mode of index $11$ in terms of prediction PSNR ($\text{PSNR} = 28.05$ dB), and (f) original block.}
	\label{figure:predictions_kodak_16_16_275}
\end{figure}

\begin{figure}
	\centering
	\begin{subfigure}{0.24\linewidth}
		\centering
		\includegraphics[width=\linewidth]{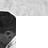}
		\caption{}
	\end{subfigure}
	\hspace{2.0mm}
	\begin{subfigure}{0.080\linewidth}
		\centering
		\includegraphics[width=\linewidth]{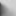}
		\caption{}
	\end{subfigure}
	\hspace{2.0mm}
	\begin{subfigure}{0.080\linewidth}
		\centering
		\includegraphics[width=\linewidth]{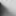}
		\caption{}
	\end{subfigure}
	\hspace{2.0mm}
	\begin{subfigure}{0.165\linewidth}
		\centering
		\vspace{5.2mm}
		\includegraphics[width=\linewidth]{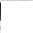}
		\caption{}
	\end{subfigure}
	\hspace{2.0mm}
	\begin{subfigure}{0.080\linewidth}
		\centering
		\includegraphics[width=\linewidth]{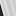}
		\caption{}
	\end{subfigure}
	\hspace{2.0mm}
	\begin{subfigure}{0.080\linewidth}
		\centering
		\includegraphics[width=\linewidth]{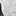}
		\caption{}
	\end{subfigure}
	\caption{Predictions of a $16 \times 16$ luminance block. The legend in Figure \ref{figure:predictions_kodak_16_16_275} is replicated here, except that (b) $\text{PSNR} = 16.06$ dB, (c) $\text{PSNR} = 13.05$ dB, and (e) the best H.265 intra prediction mode in terms of prediction PSNR has index $23$ and its prediction PSNR is equal to $21.83$ dB.}
	\label{figure:predictions_bsds_16_16_3740}
\end{figure}

Successful cases for the iterative training, i.e. with improvement of the quality of the neural network prediction over iterations, are illustrated by Figures~\ref{figure:predictions_kodak_4_4_418},~\ref{figure:predictions_kodak_8_8_193},~and~\ref{figure:predictions_kodak_16_16_275}. Failure cases are depicted in Figures~\ref{figure:predictions_kodak_4_4_688},~\ref{figure:predictions_kodak_8_8_209},~and~\ref{figure:predictions_bsds_16_16_3740}. In all cases, the neural network prediction gets shaper over iterations. Note that this indicates that the training data cleansing reaches its goal.

In Figure~\ref{figure:predictions_kodak_4_4_688}, the predicted block given by the neural network after the first iteration and the one from the best H.265 mode are all black. But, after the third iteration, the bottom-right of the predicted block returned by the neural network contains an extrapolation of the diagonal bright gray border at the top-right of the context. In Figure~\ref{figure:predictions_kodak_8_8_209}, the predicted block provided by the neural network after the first iteration and the one from the best H.265 mode are all dark gray. However, after the third iteration, we see at the bottom-right of the predicted block generated by the neural network an extrapolation of the diagonal bright gray border at the bottom of the context. Similarly, in Figure~\ref{figure:predictions_bsds_16_16_3740}, the predicted block computed by the neural network after the first iteration looks blurry whereas, after the third iteration, it includes a black triangle, which contrasts with the predicted block from the best H.265 mode. According to the last three examples, the neural network can diverge from the best H.265 mode over iterations. Note that this suggests that the iterative aspect of the training actually achieves its target. Differently, in Figures~\ref{figure:predictions_kodak_4_4_418}~and~\ref{figure:predictions_kodak_16_16_275}, the predicted block given by the neural network comes closer to the one from the best H.265 mode over iterations. This convergence might stem from the very high likelihood of the best H.265 mode given the context of the block.


\begin{figure}
	\centering
	\begin{subfigure}{0.48\linewidth}
		\centering
		\includegraphics[width=\linewidth]{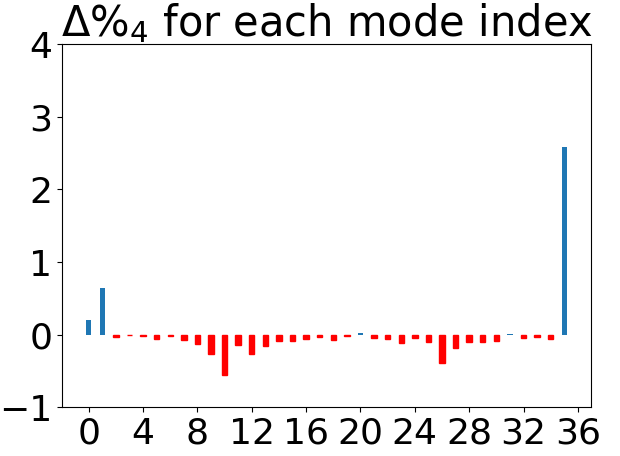}
		\caption{}
	\end{subfigure}
	\begin{subfigure}{0.48\linewidth}
		\centering
		\includegraphics[width=\linewidth]{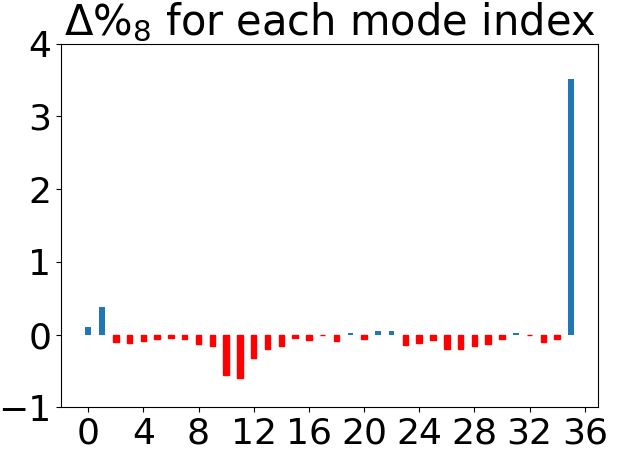}
		\caption{}
	\end{subfigure}
	\begin{subfigure}{0.48\linewidth}
		\centering
		\vspace{2.0mm}
		\includegraphics[width=\linewidth]{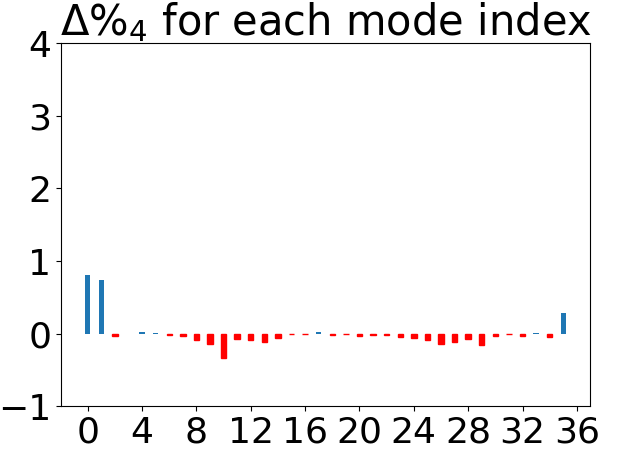}
		\caption{}
	\end{subfigure}
	\begin{subfigure}{0.48\linewidth}
		\centering
		\vspace{2.0mm}
		\includegraphics[width=\linewidth]{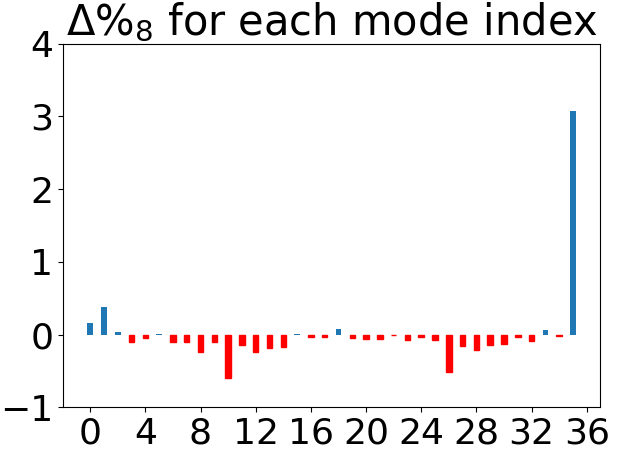}
		\caption{}
	\end{subfigure}
	\caption{Difference $\Delta\%_{w}$ between the percentage of selection an intra prediction mode in H.265-ITNN after the third training iteration and that after the first training iteration on the $w \times w$ luminance blocks returned by the image partitioning: (a) $w = 4$ and $\text{QP} = 22$, (b) $w = 8$ and $\text{QP} = 22$, (c) $w = 4$ and $\text{QP} = 27$, (d) $w = 8$ and $\text{QP} = 27$. The index of the single additional neural network-based intra prediction mode is $35$. $\Delta\%_{w}$ is computed by averaging over the H.265-ITNN encodings of the $124$ Y$\text{C}_{\text{b}}\text{C}_{\text{r}}$ images used earlier.}
	\label{figure:variations_frequencies_over_iterations}
	\vspace{-3.0mm}
\end{figure}

Given the conclusions of the last two paragraphs, in H.265-ITNN, if the context of a block contains a clear border along the frontier with this block, the single additional mode should infer from the context a likely direction of propagation. It is either the correct direction and the prediction quality improves over iterations or an incorrect direction and the prediction quality degrades over iterations. That is why, on average, the frequency of selection of the single additional mode in H.265-ITNN should increase over iterations while those of the H.265 directional modes must decrease and those of PLANAR and DC should go up. This must be valid for small blocks at small QPs since the large blocks returned by the image partitioning at large QPs sometimes exhibit complex textures, see Section \ref{subsection:generic_intra_prediction}, and the behavior of a deep predictor on large blocks with diverse textures can hardly be forecast \cite{context_encoders_feature}. The previous assumption is verified by Figure~\ref{figure:variations_frequencies_over_iterations}. Therefore, for small blocks, the single additional mode takes a growing proportion of frequency of selection away from the H.265 directional modes over iterations and, as its signalling cost is much smaller \cite{context_adaptive_neural}, bitrates are saved.

% Signalling of the neural network-based mode in H.266.
% !TeX root = iterative_training_of.tex

\section{Signalling of the neural network-based intra prediction mode in H.266} \label{section:signalling_of_the}
Section \ref{subsection:benefits_of_the} evaluates the rate-distortion performance of H.265 including the single additional neural network-based intra prediction mode. Before assessing the rate-distortion performance of H.266 with the single additional mode, a signalling of this mode in H.266 must be developed.

The first principle of the proposed signalling addresses the large number of neural networks in H.266 incurring a large memory cost. Indeed, blocks of each possible size are predicted by a different neural network belonging to the single additional mode, see Section \ref{section:global_setup}. As the H.266 partitioning can involve $25$ different block sizes, the single additional mode in H.266 should be made of $25$ neural networks. Note that the calculation of the previous figure excludes the option of splitting a luminance PB into multiple TBs because, as said in Section \ref{subsection:signalling_for_a_luminance}, the combination of the neural network intra prediction and this type of split is not allowed. This contrasts with the H.265 partitioning involving $5$ different block sizes, the single additional mode thus comprising $5$ neural networks only. As a deep neural network for intra prediction has a few million parameters, see Section \ref{subsection:benefits_of_the}, the memory cost of the parameters of $25$ neural networks in H.266 seems to be too high. To reduce this cost, only blocks of some sizes are predicted via neural networks. For each of these sizes, the pair of the height and width is inserted into the set $T \subseteq Q$. Then, for the current block to be predicted, the single additional mode is signalled if it includes the neural network predicting blocks of the current block size.

Furthermore, the single additional mode should not be signalled when the neural network prediction cannot be carried out as the context of the current block to be predicted goes out of the image bounds.
\begin{figure}
	\centering
	\includegraphics[width=\linewidth]{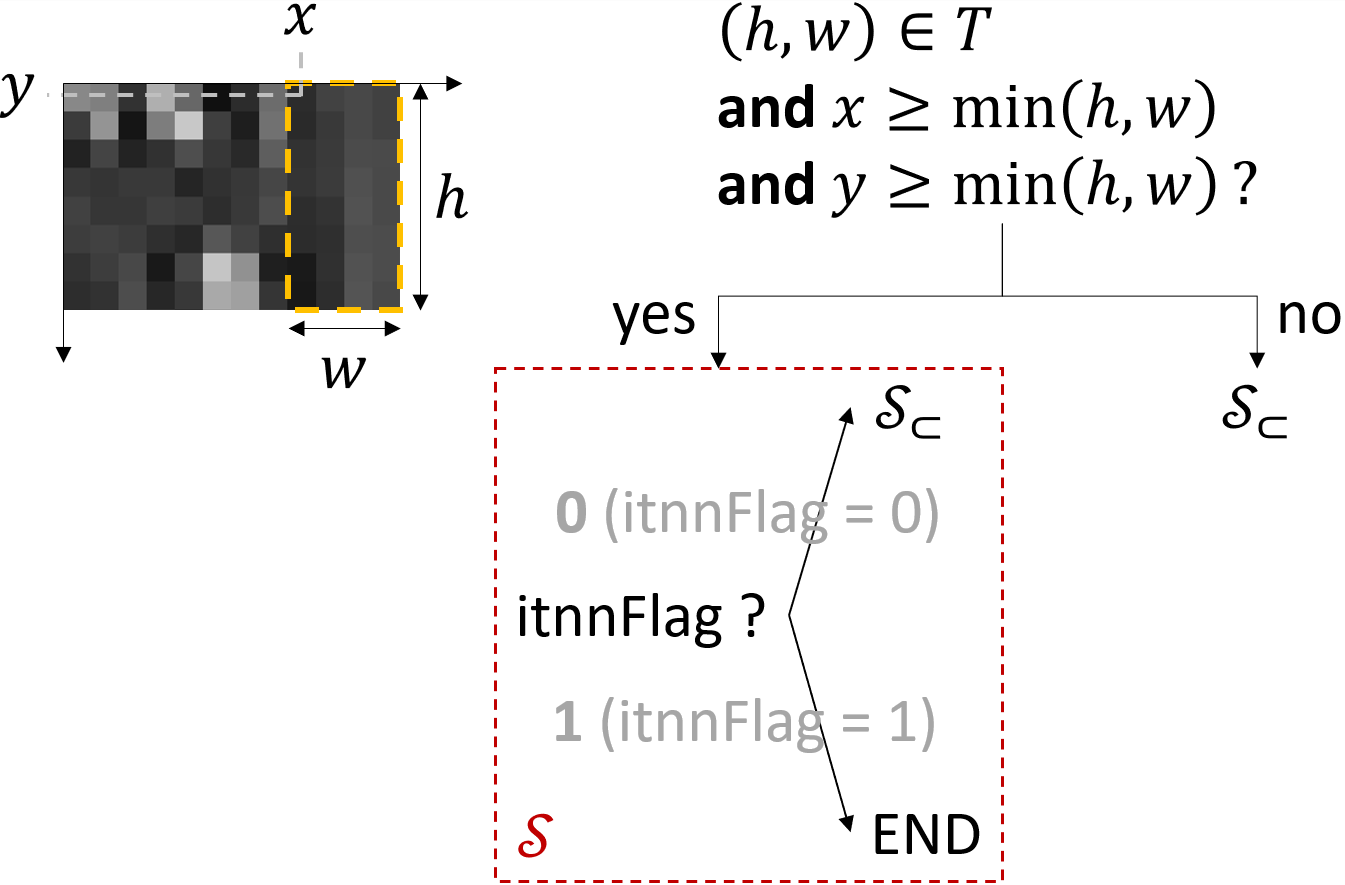}
	\caption{Choice of the intra prediction mode signalling for the current $w \times h$ luminance PB framed in orange. The first frame of \enquote{PartyScene} in $4$:$2$:$0$ is being encoded via H.266-ITNN with $\text{QP} = 37$. The coordinates of the pixel at the top-left of this PB are $x = 8$ and $y =0$. $h = 8$ and $w = 4$. The bin value of a \textit{itnnFlag} value appears in bold gray.}
	\label{figure:signalling_luminance_switch}
\end{figure}

\subsection{Signalling for a luminance PB} \label{subsection:signalling_for_a_luminance}
Given the above-mentioned two principles, for a $w \times h$ luminance PB whose top-left pixel is located at $\left( x, y \right)$ in the image, the intra prediction mode signalling $\mathcal{S}$ including the flag \textit{itnnFlag} of the single additional mode is chosen if $\left( h, w \right) \in T$ and $x \geq \text{min} \left( h, w \right)$ and $y \geq \text{min} \left( h, w \right)$. Otherwise, another intra prediction mode signalling $\mathcal{S}_{\subset}$ which does not comprise \textit{itnnFlag} applies, see Figure \ref{figure:signalling_luminance_switch}.

Moreover, as the single additional mode yields predictions of good average quality over various luminance blocks, see Figures \ref{figure:predictions_kodak_4_4_418} to \ref{figure:predictions_bsds_16_16_3740}, according to entropy coding, \textit{itnnFlag} should be placed first in $\mathcal{S}$, see Figure \ref{figure:signalling_luminance_switch}. \textit{itnnFlag} $= 1$ indicates that the single additional mode predicts the current luminance PB.

Since VTM-5.0, H.266 has featured MIP, a machine learning-based intra prediction tool, see Section \ref{subsection:intra_prediction_in}. To compare in terms of rate-distortion H.266 to H.266 with the single additional mode, called H.266-ITNN, each codec having a different machine learning-based intra prediction tool, MIP is removed from H.266-ITNN. Therefore, $\mathcal{S}_{\subset}$ denotes the H.266 intra prediction mode signalling for the current luminance PB without the MIP signalling. Note that, given Figure \ref{figure:signalling_luminance_switch}, the neural network-based intra prediction cannot be combined with Intra Sub-Partitions (ISP) \cite{intra_sub_partitions}, which allows to partition the current luminance PB into multiple TBs.

\subsection{Signalling for a chrominance PB} \label{subsection:signalling_for_a_chrominance}
As with luminance, the single additional mode gives predictions of good average quality over various chrominance blocks, implying that its signalling cost must be low for chrominance too. As the flag of the Direct Mode (DM) comes first in the intra prediction mode signalling for a chrominance PB in VTM-5.0 \cite{versatile_video_coding_draft_5}, we allow the neural network intra prediction of a chrominance PB via DM. Note that, since VTM-6.0, the DM flag has no longer been first in the intra prediction mode signalling for a chrominance PB since the Cross-Component Linear Model flag has been placed before it \cite{versatile_video_coding_draft_6}.

But, the use of the single additional mode via DM is restricted by the two principles laid down at the beginning of Section \ref{section:signalling_of_the}. To incorporate these constraints, for a given $w \times h$ chrominance PB whose top-left pixel is at $\left( x, y \right)$ in the image, if the luminance PB collocated with this chrominance PB is predicted by the single additional mode, DM becomes the single additional mode if $\left( h, w \right) \in T$ and $x \geq \text{min} \left( h, w \right)$ and $y \geq \text{min} \left( h, w \right)$. Otherwise, DM is set to PLANAR.

\subsection{Sizes of blocks predicted by the neural network-based mode} \label{subsection:set_of_the}
Now, only the set $T$ of pairs of the height and width of the blocks that can be predicted by the single additional mode remains to be defined. For non-square blocks, when the single additional mode includes neural networks predicting relatively small blocks, e.g. $4 \times 8$ blocks, and does not contain those predicting the large ones, e.g. $8 \times 32$ blocks, the rate-distortion performance is better compared to the other way round, see Section \ref{subsection:comparison_for_h266}. By favoring the relatively small non-square blocks over the large ones, $T = \left\{ (4, 4), (4, 8), (8, 4), (8, 8), (16, 16), (4, 16), (16, 4), (8, 16), \right.$ $\left. (16, 8), (32, 32), (64, 64) \right\}$. Therefore, the single additional mode comprises $11$ neural networks in H.266-ITNN.

% Comparison with the state-of-the-art.
% !TeX root = iterative_training_of.tex

\section{Comparison with the state-of-the-art} \label{section:comparison_with_the}
Now that the iterative training of neural networks for intra prediction is explained, see Section \ref{section:iterative_training_of}, and the single additional mode involving the trained neural networks can be integrated into both H.265 and H.266, see Section \ref{section:signalling_of_the}, the two codecs with the single additional mode, namely H.265-ITNN and H.266-ITNN, can be compared against the state-of-the-art.

\setlength{\tabcolsep}{8pt}
\begin{table*}
	\caption{BD-rate reductions in $\%$ of H.265-ITNN with $l = 3$, PS-RNN \cite{progressive_spatial_recurrent}, IPFCN-D \cite{fully_connected_network}, and H.265 with the generative adversarial neural network-based intra prediction mode in \cite{generative_adversarial_network}. As H.265-ITNN is built from HM-16.15, the anchor of H.265-ITNN is HM-16.15. For similar reasons, the anchors of PS-RNN, IPFCN-D, and H.265 with the generative adversarial neural network-based mode are HM-16.15, HM-16.9, and HM-16.17 respectively. Only the first frame of each video sequence is considered. All BD-rate reductions have one decimal digit, similarly to \cite{fully_connected_network}. The largest BD-rate reduction in absolute value for the luminance channel is shown in bold. Note that, using the current experimental setups, HM-16.15 and HM-16.17 yield the same rate-distortion scores. Therefore, HM-16.15 and HM-16.17 are two equivalent anchors.}
	\centering
	\begin{tabular}{llcccccccccccc}
		\hline
		\multicolumn{2}{l}{\multirow{2}{*}{Video sequence}} & \multicolumn{3}{c}{our H.265-ITNN} & \multicolumn{3}{c}{PS-RNN \cite{progressive_spatial_recurrent}} & \multicolumn{3}{c}{IPFCN-D \cite{fully_connected_network}} & \multicolumn{3}{c}{Generative method \cite{generative_adversarial_network}}\\ \cline{3-14} & & Y & $\text{C}_{\text{b}}$ & $\text{C}_{\text{r}}$ & Y & $\text{C}_{\text{b}}$ & $\text{C}_{\text{r}}$ & Y & $\text{C}_{\text{b}}$ & $\text{C}_{\text{r}}$ & Y & $\text{C}_{\text{b}}$ & $\text{C}_{\text{r}}$\\
		\hline
		\multirow{5}{*}{B} & BasketballDrive & $\mathbf{-8.5}$ & $-9.7$ & $-9.5$ & $-1.4$ & $-1.1$ & $-1.4$ & $-3.6$ & $-2.9$ & $-2.7$ & $-0.9$ & $0.2$ & $-1.5$\\ & BQTerrace & $-4.1$ & $-4.4$ & $-4.0$ & $-2.4$ & $-0.5$ & $-0.5$ & $-2.1$ & $-1.3$ & $-0.5$ & $\mathbf{-7.2}$ & $-8.6$ & $-8.6$\\ & Cactus & $\mathbf{-4.2}$ & $-4.1$ & $-4.2$ & $-2.3$ & $-1.5$ & $-0.9$ & $-3.2$ & $-1.8$ & $-1.5$ & $-2.3$ & $-2.2$ & $-2.8$\\ & Kimono & $\mathbf{-3.2}$ & $-3.8$ & $-2.9$ & $-1.2$ & $-0.9$ & $-0.9$ & $-3.1$ & $-2.1$ & $-1.5$ & $-0.0$ & $-0.7$ & $-1.0$\\ & ParkScene & $-2.3$ & $-4.0$ & $-3.4$ & $-2.7$ & $-1.6$ & $-1.3$ & $\mathbf{-3.6}$ & $-2.2$ & $-2.4$ & $-1.0$ & $-1.6$ & $-1.0$\\
		\hline
		\multirow{4}{*}{C} & BasketballDrill & $-5.6$ & $-7.9$ & $-7.1$ & $-1.6$ & $-0.3$ & $-1.5$ & $-1.5$ & $-3.3$ & $-2.2$ & $\mathbf{-11.1}$ & $-10.3$ & $-11.6$\\ & BQMall & $-4.1$ & $-5.0$ & $-5.5$ & $-3.0$ & $-1.4$ & $0.0$ & $-2.2$ & $-1.9$ & $-1.0$ & $\mathbf{-17.3}$ & $-18.6$ & $-20.4$\\ & PartyScene & $-3.0$ & $-3.0$ & $-2.8$ & $-2.5$ & $-2.2$ & $-2.2$ & $-1.6$ & $-1.2$ & $-0.1$ & $\mathbf{-7.9}$ & $-9.3$ & $-9.4$\\ & RaceHorses & $ \mathbf{-3.8}$ & $-4.1$ & $-3.2$ & $-2.3$ & $-1.8$ & $-1.0$ & $-3.2$ & $-1.9$ & $-2.8$ & $-0.8$ & $-0.9$ & $-1.7$\\
		\hline
		\multirow{4}{*}{D} & BasketballPass & $\mathbf{-3.8}$ & $-4.3$ & $-4.1$ & $-2.1$ & $-1.9$ & $-0.5$ & $-1.2$ & $-0.3$ & $1.1$ & $-0.1$ & $0.6$ & $0.6$\\ & BlowingBubbles & $\mathbf{-3.0}$ & $-5.2$ & $-4.3$ & $-2.7$ & $-2.2$ & $0.5$ & $-1.9$ & $-2.8$ & $-3.5$ & $-1.8$ & $-5.3$ & $-5.0$\\ & BQSquare & $-3.3$ & $-2.0$ & $-1.9$ & $-2.1$ & $-0.6$ & $1.8$ & $-0.9$ & $-0.1$ & $-2.8$ & $\mathbf{-10.0}$ & $-12.3$ & $-12.3$\\ & RaceHorses & $\mathbf{-5.3}$ & $-5.3$ & $-6.3$ & $-3.5$ & $-3.3$ & $-1.8$ & $-3.2$ & $-2.6$ & $-2.8$ & $-0.8$ & $-1.3$ & $-1.6$\\
		\hline
		\multicolumn{2}{l}{Mean} & $-4.2$ & $-4.8$ & $-4.5$ & $-2.3$ & $-1.5$ & $-0.7$ & $-2.4$ & $-1.9$ & $-1.7$ & $\mathbf{-4.7}$ & $-5.4$ & $-5.9$\\
		\hline 
	\end{tabular}
	\label{table:comparison_with_the_265}
	\vspace{-2.0mm}
\end{table*}

\subsection{Comparison for H.265} \label{subsection:comparison_for_h265}
In the literature on the enhancement of the H.265 intra prediction via neural networks, Progressive Spatial Recurrent Neural Network (PS-RNN) \cite{progressive_spatial_recurrent}, Intra Prediction Fully-Connected Networks (IPFCN) \cite{fully_connected_network}, the generative adversarial neural network-based intra prediction in \cite{generative_adversarial_network}, and the Neural Network-based intra prediction modes (NN-modes) \cite{intra_picture_prediction}, see Section \ref{subsection:neural_network_based}, stand out as the four benchmarked approaches.

\subsubsection{Comparison with \cite{progressive_spatial_recurrent, fully_connected_network, generative_adversarial_network}} \label{subsubsection:comparison_state_0}
The experimental setup shared by \cite{progressive_spatial_recurrent, fully_connected_network, generative_adversarial_network} is reproduced. Specifically, only the first frame of each video sequence from the classes B, C, and D of the CTC \cite{common_test_conditions} is considered. The rate-distortion performance of H.265 including a neural network-based intra prediction mode is calculated via the Bjontegaard metric of this modified version of H.265 with respect to H.265 with $\text{QP} \in \left\{ 22, 27, 32, 37 \right\}$. All intra is used as configuration.

IPFCN comes in two variants, namely Intra Prediction Fully-Connected Networks Single (IPFCN-S) and Intra Prediction Fully-Connected Networks Dual (IPFCN-D), the latter featuring an enhanced training process. Indeed, in IPFCN-S, four fully-connected networks are designed to predict blocks of sizes $4 \times 4$, $8 \times 8$, $16 \times 16$, and $32 \times 32$ respectively. Then, each neural network is trained on pairs of a luminance block returned by the H.265 partitioning of images and its context. Finally, the four trained neural networks are aggregated into a neural network-based mode in H.265. In contrast, in IPFCN-D, the above-mentioned set of fully-connected networks is duplicated into two sets. Then, the different pairs of a luminance block given by the H.265 partitioning of images and its context are clustered into two groups, the first one gathering the blocks predicted via either PLANAR or DC and the second one containing those predicted via a directional mode. Each of the two sets of neural networks is trained on a different cluster. In the end, the two sets of trained neural networks form the neural network-based mode in H.265. As IPFCN-D outperforms IPFCN-S in terms of rate-distortion \cite{fully_connected_network}, IPFCN-D is picked for comparison.

Table \ref{table:comparison_with_the_265} shows that, for the luminance channel, H.265-ITNN using three training iterations yields $-1.8\%$ of additional mean BD-rate reduction compared to IPFCN-D. With respect to PS-RNN, the additional mean BD-rate reduction reaches $-1.9\%$. The approach in \cite{generative_adversarial_network} gives $-0.5\%$ of additional mean BD-rate reduction with respect to H.265-ITNN. But, the complexity of the prediction mechanism in \cite{generative_adversarial_network} far exceeds those of H.265-ITNN, PS-RNN, and IPFCN-D. Indeed, in \cite{generative_adversarial_network}, for a given $w \times w$ PB, $w \in \left\{ 4, 8, 16, 32, 64 \right\}$, a 17-layer convolutional neural network predicts the $64 \times 64$ block whose top-left pixel matches the top-left pixel of the $w \times w$ PB, and the prediction of the $w \times w$ PB is obtained by cropping the $w \times w$ area at the top-left of the $64 \times 64$ neural network prediction. In contrast, in H.265-ITNN, PS-RNN, and IPFCN-D, a smaller neural network directly returns a $w \times w$ prediction of a given $w \times w$ PB. This makes a sensitive difference in decoder run time, see Section \ref{subsection:complexity}.

\subsubsection{Comparison with \cite{intra_picture_prediction}}
As several experimental setups in \cite{intra_picture_prediction} differ from those shared by \cite{progressive_spatial_recurrent, fully_connected_network, generative_adversarial_network}, another experiment following the experimental setups in \cite{intra_picture_prediction} compares H.265-ITNN to NN-modes. This time, all the frames of each video sequence from the classes A, B, C, and D used in \cite{intra_picture_prediction} are considered. As in Section \ref{subsubsection:comparison_state_0}, $\text{QP} \in \left\{ 22, 27, 32, 37 \right\}$ and the configuration is all intra.

For the luminance channel, H.265-ITNN using three training iterations yields $-1.25\%$ of additional mean BD-rate reduction with respect to NN-modes, see Table \ref{table:comparison_intra_picture_prediction}. Note that, as NN-modes refers to VTM-1.0 with multiple neural network-based intra prediction modes, its anchor is VTM-1.0. Although H.265-ITNN and NN-modes do not have the same anchor, their comparison is still valid in regard to intra prediction for two reasons. Firstly, only the partitioning changes from HM-16.15 to VTM-1.0. Indeed, VTM-1.0 amounts to HM-16.15 with QuadTree plus Binary Tree and Ternary Tree \cite{quadtree_plus_binary, highly_flexible_coding}. Secondly, given that each mode in NN-modes predicts blocks of each size via a different neural network, the performance of the neural network-based intra prediction should barely be affected by the modifications of partitioning.
\setlength{\tabcolsep}{2.8pt}
\begin{table}
	\vspace{-2.0mm}
	\caption{BD-rate reductions in $\%$ of H.265-ITNN with $l = 3$ and NN-modes \cite{intra_picture_prediction}. For H.265-ITNN, the anchor is H.265 (HM-16.15). For NN-modes, the anchor is VTM-1.0. All the frames of each video sequence are considered. All BD-rate reductions have two decimal digits, similarly to \cite{intra_picture_prediction}. The largest BD-rate reduction in absolute value for the luminance channel is shown in bold.}
	\centering
	\begin{tabular}{llcccccc}
		\hline
		\multicolumn{2}{l}{\multirow{2}{*}{Video sequence}} & \multicolumn{3}{c}{our H.265-ITNN} & \multicolumn{3}{c}{NN-modes \cite{intra_picture_prediction}}\\ \cline{3-8} & & Y & $\text{C}_{\text{b}}$ & $\text{C}_{\text{r}}$ & Y & $\text{C}_{\text{b}}$ & $\text{C}_{\text{r}}$\\
		\hline
		\multirow{6}{*}{A} & Tango2 & $\mathbf{-6.81}$ & $-6.77$ & $-9.42$ & $-5.20$ & $-4.31$ & $-3.42$\\ & FoodMarket4 & $\mathbf{-13.42}$ & $-11.90$ & $-12.35$ & $-5.67$ & $-2.90$ & $-3.06$\\ & CampfireParty & $\mathbf{-2.63}$ & $-3.70$ & $-3.74$ & $-1.44$ & $-0.88$ & $-1.29$\\ & CatRobot1 & $\mathbf{-4.79}$ & $-5.37$ & $-5.80$ & $-3.66$ & $-2.69$ & $-1.96$\\ & DaylightRoad2 & $\mathbf{-4.52}$ & $-8.07$ & $-8.16$ & $-4.01$ & $-1.60$ & $-2.47$\\ & ParkRunning3 & $\mathbf{-2.49}$ & $-2.99$ & $-2.95$ & $-1.93$ & $-1.81$ & $-2.24$\\
		\hline
		\multirow{5}{*}{B} & MarketPlace & $-2.50$ & $-3.75$ & $-4.61$ & $\mathbf{-3.11}$ & $-1.33$ & $-0.48$\\ & RitualDance & $-5.35$ & $-5.59$ & $-7.24$ & $\mathbf{-5.49}$ & $-2.89$ & $-2.68$\\ & BasketballDrive & $\mathbf{-6.90}$ & $-8.89$ & $-9.19$ & $-2.92$ & $-1.95$ & $-2.00$\\ & BQTerrace & $\mathbf{-3.85}$ & $-4.76$ & $-4.24$ & $-2.60$ & $-0.60$ & $0.10$\\ & Cactus & $\mathbf{-4.09}$ & $-4.45$ & $-4.37$ & $-3.88$ & $-2.17$ & $-1.99$\\
		\hline
		\multirow{4}{*}{C} & BasketballDrill & $\mathbf{-6.13}$ & $-5.58$ & $-5.89$ & $-2.51$ & $-2.32$ & $-2.61$\\ & BQMall & $-4.32$ & $-4.17$ & $-5.52$ & $\mathbf{-4.40}$ & $-2.47$ & $-2.29$\\ & PartyScene & $\mathbf{-2.94}$ & $-3.19$ & $-2.87$ & $-2.89$ & $-1.56$ & $-1.62$\\ & RaceHorses & $\mathbf{-3.36}$ & $-4.33$ & $-3.21$ & $-2.78$ & $-1.63$ & $-1.88$\\
		\hline
		\multirow{4}{*}{D} & BasketballPass & $\mathbf{-4.12}$ & $-4.23$ & $-3.90$ & $-3.43$ & $-2.22$ & $-2.52$\\ & BlowingBubbles & $\mathbf{-3.13}$ & $-4.24$ & $-4.14$ & $-2.71$ & $-1.63$ & $-2.00$\\ & BQSquare & $\mathbf{-3.33}$ & $-3.59$ & $-2.43$ & $-2.84$ & $-0.79$ & $-1.05$\\ & RaceHorses & $\mathbf{-4.24}$ & $-5.12$ & $-7.35$ & $-3.74$ & $-2.43$ & $-2.00$\\
		\hline
		\multicolumn{2}{l}{Mean} & $\mathbf{-4.68}$ & $-5.30$ & $-5.65$ & $-3.43$ & $-2.01$ & $-1.97$\\
		\hline 
	\end{tabular}
	\label{table:comparison_intra_picture_prediction}
\end{table}

\subsection{Comparison for H.266} \label{subsection:comparison_for_h266}
As H.266 includes MIP since VTM-5.0, the evaluation of H.266-ITNN, which is built from VTM-5.0 without MIP, with respect to VTM-5.0 establishes a direct comparison between our single additional mode and MIP. The experimental protocol in this section takes the protocol from Section \ref{subsubsection:comparison_state_0}, except that eight video sequences from the class A of the CTC are added in order to diversify the test data.

Due to the increase of encoder run time from HM to VTM, the commonly used deep neural networks for intra prediction cannot currently be trained in reasonable time via the proposed iterative training with H.266 as codec. Indeed, even though the encoder run time of H.265-ITNN is $50$ times larger than that of HM-16.15, see section \ref{subsection:complexity}, the encoding of $\Gamma$ via H.265-ITNN does not exceed $4$ days using $400$ cores. However, as the ratio between the encoder run time of VTM-5.0 and that of HM-16.15 is nearly $18$, the encoding of $\Gamma$ via H.266 with the single additional mode seems to be too long. Two solutions could be the optimization of the neural network architectures for faster inference \cite{efficient_and_accurate, nips_pruning_networks} and the implementation of heuristics for VTM encoder acceleration. But, these heuristics have not been developed yet since the H.266 standard is not finalized yet. For now, the models trained iteratively with H.265 as codec can be inserted into H.266 at test time. Therefore, the following experiments involve two versions of H.266-ITNN. For the \enquote{true} H.266-ITNN, blocks of sizes $4 \times 4$, $8 \times 8$, $16 \times 16$, and $32 \times 32$ are predicted by the four neural networks trained iteratively with H.265 as codec and $l = 3$, the blocks of sizes $8 \times 4$, $4 \times 8$, $16 \times 4$, $4 \times 16$, $16 \times 8$, $8 \times 16$, and $64 \times 64$ being predicted by neural networks trained as in \cite{context_adaptive_neural}, c.f. Section \ref{subsection:neural_network_based}. For the second version, called H.266-STNN, any block whose pair of height and width belongs to $T$ is predicted by a neural network trained as in \cite{context_adaptive_neural}. STNN stands for Simply Trained Neural Networks.
\begin{figure*}
	\centering
	\begin{subfigure}{0.24\linewidth}
		\centering
		\includegraphics[width=\linewidth]{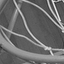}
		\caption{}
	\end{subfigure}
	\begin{subfigure}{0.24\linewidth}
		\centering
		\includegraphics[width=\linewidth]{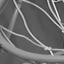}
		\caption{}
	\end{subfigure}
	\begin{subfigure}{0.24\linewidth}
		\centering
		\includegraphics[width=\linewidth]{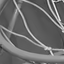}
		\caption{}
	\end{subfigure}
	\begin{subfigure}{0.24\linewidth}
		\centering
		\includegraphics[width=\linewidth]{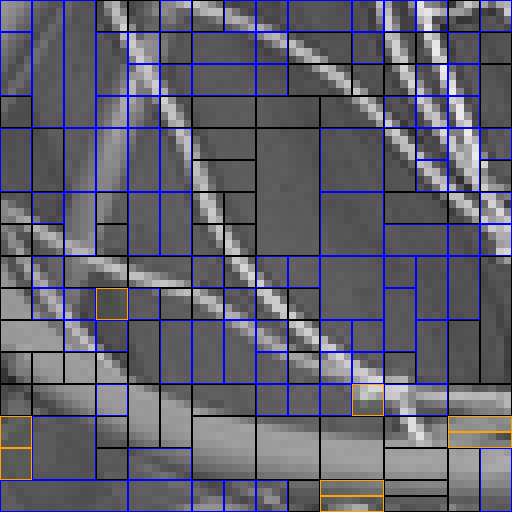}
		\caption{}
	\end{subfigure}
	\caption{Comparison of (a) the $64 \times 64$ block whose top-left pixel is at $\left( 64, 64 \right)$ in the luminance channel of \enquote{BasketballDrill}, (b) its reconstruction via VTM-5.0, (c) its reconstruction via H.266-ITNN, and (d) its partitioning via H.266-ITNN. $\text{QP} = 27$. In (d), the mapping from the color of the frame of the block to the intra prediction mode of the block is \enquote{orange} $\rightarrow$ either PLANAR or DC, \enquote{black} $\rightarrow$ directional mode, and \enquote{blue} $\rightarrow$ neural network-based mode. For the entire luminance channel, for VTM-5.0, rate = $0.310$ bpp and PSNR $= 39.503$ dB. For H.266-ITNN, rate = $0.305$ bpp and PSNR $= 39.491$ dB.}
	\label{figure:visualization_h266_itnn}
	\vspace{-3.0mm}
\end{figure*}
\setlength{\tabcolsep}{3.5pt}
\begin{table}
	\caption{BD-rate reductions in $\%$ of H.266-STNN and H.266-ITNN. The anchor is VTM-5.0. Only the first frame of each video sequence is considered. The largest BD-rate reduction in absolute value for the luminance channel is displayed in bold.}
	\centering
	\begin{tabular}{llcccccc}
		\hline
		\multicolumn{2}{l}{\multirow{2}{*}{Video sequence}} & \multicolumn{3}{c}{H.266-STNN} & \multicolumn{3}{c}{H.266-ITNN}\\ \cline{3-8} & & Y & $\text{C}_{\text{b}}$ & $\text{C}_{\text{r}}$ & Y & $\text{C}_{\text{b}}$ & $\text{C}_{\text{r}}$\\
		\hline
		\multirow{8}{*}{A} & CampfireParty & $-1.16$ & $-0.53$ & $-0.30$ & $\mathbf{-1.29}$ & $-0.61$ & $-0.25$\\ & DaylightRoad2 & $-1.76$ & $0.57$ & $-2.32$ & $\mathbf{-2.04}$ & $-1.00$ & $-1.58$\\ & Drums2 & $-1.44$ & $-1.64$ & $-1.03$ & $\mathbf{-1.63}$ & $-1.76$ & $-1.10$\\ & Tango2 & $-2.54$ & $-2.35$ & $-1.96$ & $\mathbf{-2.95}$ & $-2.09$ & $-3.03$\\ & ToddlerFountain2 & $-1.45$ & $-2.25$ & $-1.00$ & $\mathbf{-1.56}$ & $-2.07$ & $-1.31$\\  & TrafficFlow & $-1.09$ & $0.83$ & $0.20$ & $\mathbf{-1.45}$ & $1.57$ & $0.58$\\ & PeopleOnStreet & $-2.68$ & $-3.27$ & $-2.88$ & $\mathbf{-2.84}$ & $-2.90$ & $-3.15$\\  & Traffic & $-1.87$ & $-1.52$ & $-1.30$ & $\mathbf{-1.97}$ & $-1.62$ & $-1.90$\\
		\hline
		\multirow{5}{*}{B} & BasketballDrive & $-1.84$ & $-1.75$ & $-2.15$ & $\mathbf{-2.03}$ & $-1.34$ & $-2.22$\\ & BQTerrace & $-1.58$ & $-0.51$ & $-0.65$ & $\mathbf{-1.75}$ & $-0.22$ & $-1.67$\\ & Cactus & $-1.76$ & $-0.81$ & $-1.48$ & $\mathbf{-2.00}$ & $-1.38$ & $-1.65$\\ & Kimono & $-1.38$ & $-2.56$ & $-1.94$ & $\mathbf{-1.48}$ & $-2.72$ & $-2.51$\\ & ParkScene & $-1.18$ & $-1.65$ & $-2.08$ & $\mathbf{-1.22}$ & $-0.71$ & $0.45$\\
		\hline
		\multirow{4}{*}{C} & BasketballDrill & $-1.06$ & $0.65$ & $-0.41$ & $\mathbf{-1.26}$ & $0.50$ & $-1.94$\\ & BQMall & $-2.21$ & $-1.80$ & $-1.89$ & $\mathbf{-2.60}$ & $-1.14$ & $-1.11$\\ & PartyScene & $-1.96$ & $-3.16$ & $-0.88$ & $\mathbf{-2.15}$ & $-2.88$ & $-1.32$\\ & RaceHorses & $-1.84$ & $-0.34$ & $-1.86$ & $\mathbf{-1.90}$ & $-1.18$ & $0.09$\\
		\hline
		\multirow{4}{*}{D} & BasketballPass & $-1.19$ & $-3.41$ & $-4.13$ & $\mathbf{-1.73}$ & $-0.46$ & $-3.02$\\ & BlowingBubbles & $-1.80$ & $-0.64$ & $-0.10$ & $\mathbf{-2.05}$ & $-1.65$ & $1.00$\\ & BQSquare & $-1.86$ & $0.96$ & $-2.54$ & $\mathbf{-2.13}$ & $1.63$ & $-2.98$\\ & RaceHorses & $-2.00$ & $-3.67$ & $-1.15$ & $\mathbf{-2.27}$ & $-1.90$ & $-1.88$\\
		\hline
		\multicolumn{2}{l}{Mean} & $-1.70$ & $-1.37$ & $-1.52$ & $\mathbf{-1.92}$ & $-1.14$ & $-1.45$\\
		\hline 
	\end{tabular}
	\label{table:comparison_with_the_h266}
\end{table}
\setlength{\tabcolsep}{7.1pt}
\begin{table}
	\caption{mean BD-rate reduction per class in $\%$ of H.266-ITNN and VTM-5.0 with MIP on. The anchor is VTM-5.0 with MIP off. Only the first frame of each of the video sequences in Table \ref{table:comparison_with_the_h266} is considered. The largest mean BD-rate reduction in absolute value for the luminance channel is written in bold. \enquote{Mean} refers to an average over video sequences, not over classes.}
	\centering
	\begin{tabular}{lcccccc}
		\hline
		\multirow{2}{*}{Class} & \multicolumn{3}{c}{H.266-ITNN} & \multicolumn{3}{c}{VTM-5.0 with MIP on}\\ \cline{2-7} & Y & $\text{C}_{\text{b}}$ & $\text{C}_{\text{r}}$ & Y & $\text{C}_{\text{b}}$ & $\text{C}_{\text{r}}$\\
		\hline
		A & $\textbf{-2.36}$ & $-1.35$ & $-1.63$ & $-0.37$ & $-0.05$ & $-0.16$\\
		B & $\textbf{-2.03}$ & $-1.32$ & $-1.18$ & $-0.34$ & $-0.05$ & $0.33$\\
		C & $\textbf{-2.26}$ & $-1.81$ & $-1.83$ & $-0.28$ & $-0.65$ & $-0.75$\\
		D & $\textbf{-2.49}$ & $-1.23$ & $-1.49$ & $-0.41$ & $-0.56$ & $0.19$\\
		\hline
		Mean & $\textbf{-2.28}$ & $-1.41$ & $-1.54$ & $-0.35$ & $-0.26$ & $-0.09$\\
		\hline 
	\end{tabular}
	\label{table:comparison_anchor_h266_without_mip}
\end{table}

H.266-ITNN gives $-0.22\%$ of additional mean BD-rate reduction with respect to H.266-STNN, see Table \ref{table:comparison_with_the_h266}. Thus, a noticeable improvement of the rate-distortion performance of H.266 with the single additional mode can be attributed to the iterative training whereas only a relatively small fraction of the blocks are actually predicted by iteratively trained neural networks. By extending the iterative training to the $11$ neural networks in H.266-ITNN, a much larger increase in mean BD-rate reduction can be expected. Moreover, Table \ref{table:comparison_with_the_h266} reports a mean BD-rate reduction of $-1.92\%$ for H.266-ITNN. This experimental result shows the advantage of the single additional deep neural network-based mode over the multiple MIP modes. A visualization of reconstruction via H.266-ITNN is displayed in Figure \ref{figure:visualization_h266_itnn}.

To assess the net benefit of the single additional mode to H.266, the anchor can now be set to VTM-5.0 with MIP off. The mean BD-rate reduction of H.266-ITNN becomes $-2.28\%$, see Table \ref{table:comparison_anchor_h266_without_mip}.

Finally, to justify the choice of $T$ for H.265-ITNN in Section \ref{subsection:set_of_the}, H.266-ITNN is compared to H.266-ITNN-large for which $\left( 4, 8 \right)$ and $\left( 8, 4 \right)$ are replaced by $\left( 8, 32 \right)$ and $\left( 32, 8 \right)$ in $T$, see Table \ref{table:comparison_choice_blocks}. The drop in mean DB-rate reduction is significant when the single additional mode neglects several small blocks and predicts more large blocks.
\begin{table}
	\caption{mean BD-rate reduction per class in $\%$ of H.266-ITNN and H.266-ITNN-large. The anchor is VTM-5.0. Only the first frame of each of the video sequences in Table \ref{table:comparison_with_the_h266} is considered. The largest mean BD-rate reduction in absolute value for the luminance channel is written in bold. \enquote{Mean} refers to an average over video sequences, not over classes.}
	\centering
	\begin{tabular}{lcccccc}
		\hline
		\multirow{2}{*}{Class} & \multicolumn{3}{c}{H.266-ITNN} & \multicolumn{3}{c}{H.266-ITNN-large}\\ \cline{2-7} & Y & $\text{C}_{\text{b}}$ & $\text{C}_{\text{r}}$ & Y & $\text{C}_{\text{b}}$ & $\text{C}_{\text{r}}$\\
		\hline
		A & $\textbf{-1.97}$ & $-1.31$ & $-1.47$ & $-1.68$ & $-0.76$ & $-1.09$\\
		B & $\textbf{-1.70}$ & $-1.27$ & $-1.52$ & $-1.39$ & $-0.67$ & $-1.27$\\
		C & $\textbf{-1.98}$ & $-1.17$ & $-1.07$ & $-1.86$ & $-0.87$ & $-1.00$\\
		D & $\textbf{-2.04}$ & $-0.60$ & $-1.72$ & $-1.75$ & $-0.61$ & $-1.69$\\
		\hline
		Mean & $\textbf{-1.92}$ & $-1.14$ & $-1.45$ & $-1.68$ & $-0.78$ & $-1.27$\\
		\hline
	\end{tabular}
	\label{table:comparison_choice_blocks}
\end{table}

\subsection{Complexity} \label{subsection:complexity}
Table \ref{table:comparison_complexity} compares the average encoder run times and decoder run times of H.265-ITNN, PS-RNN \cite{progressive_spatial_recurrent}, IPFCN-D \cite{fully_connected_network}, and the generative method in \cite{generative_adversarial_network} on CPU only. A Intel Xeon CPU E5-2690 is used to compute the average run times of H.265-ITNN. The authors in \cite{intra_picture_prediction} report that, when NN-modes is compared to VTM-1.0, the average encoder run time reaches $284\%$ and the average decoder run time equals $147\%$. Therefore, the encoder and decoder slowdowns caused by the integration of the multiple neural network-based intra prediction modes in \cite{intra_picture_prediction} into a HEVC-like codec are much smaller than those reported in \cite{progressive_spatial_recurrent, fully_connected_network}, \cite{generative_adversarial_network}, and in the approach in the current paper. This must arise from the relatively small neural network architectures used in \cite{intra_picture_prediction}. Moreover, the average decoder run time of \cite{generative_adversarial_network} reaches $526400\%$. Thus, the average decoder run time of the generative method in \cite{generative_adversarial_network} exceeds those of H.265-ITNN, PS-RNN, and IPFCN-D by an order of magnitude. This must be explained by the complexity of the prediction mechanism in \cite{generative_adversarial_network}, see Section \ref{subsubsection:comparison_state_0}.

The drop of encoder run time from H.265-ITNN to H.266-ITNN in Table \ref{table:comparison_complexity} is due to the implementation. Indeed, in H.266-ITNN, as the neural network-based intra prediction cannot be coupled to ISP, see Section \ref{subsection:signalling_for_a_luminance}, the neural network prediction of a given luminance PB tested by the encoder can be computed only once.
\setlength{\tabcolsep}{8pt}
\begin{table*}
	\caption{Average encoder run times and decoder run times on CPU only. The anchors of H.265-ITNN, PS-RNN \cite{progressive_spatial_recurrent}, IPFCN-D \cite{fully_connected_network}, and the generative method in \cite{generative_adversarial_network} are HM-16.15, HM-16.15, HM-16.9, and HM-16.17 respectively. The anchor of H.266-ITNN is VTM-5.0. $100\%$ means that the tested codec and its anchor have the same run times.}
	\centering
	\begin{tabular}{lccccc}
		\hline
		\multirow{2}{*}{} & \multicolumn{4}{c}{H.265-based} & VTM-5.0-based\\ \cline{2-6} & H.265-ITNN & PS-RNN \cite{progressive_spatial_recurrent} & IPFCN-D \cite{fully_connected_network} & Generative method \cite{generative_adversarial_network} & H.266-ITNN\\
		\hline
		Encoding & $4917\%$ & $-$ & $9147\%$ & $14900\%$ & $987\%$\\
		Decoding & $17396\%$ & $20664\%$ & $23011\%$ & $526400\%$ & $15171\%$\\
		\hline 
	\end{tabular}
	\label{table:comparison_complexity}
\end{table*}

% Conclusion and perspectives.
% !TeX root = iterative_training_of.tex

\section{Conclusion} \label{section:conclusion}
This paper has introduced an iterative training of neural networks for intra prediction in a block-based codec. At each iteration, the neural networks become more beneficial intra predictors for this codec as the training sets are filled with pairs of a block and its context typically found in the codec image partitioning and for which the neural network intra prediction can outdo the codec intra prediction. When inserted into both H.265 and H.266, the iteratively trained neural networks bring significant improvements in terms of rate-distortion.

\bibliographystyle{IEEEbib}
\bibliography{iterative_training_of}

\begin{thebibliography}{10}

\bibitem{video_inpainting_of}
Alasdair Newson, Andr{\'e}s Almansa, Matthieu Fradet, Yann Gousseau, and
  Patrick P{\'e}rez,
\newblock ``Video inpainting of complex scenes,''
\newblock {\em SIAM Journal on Imaging Sciences}, vol. 7, no. 4, pp.
  1993--2019, October 2014.

\bibitem{multiple_linear_regression}
Zhaobin Zhang, Yue Li, Li~Li, Zhu Li, and Shan Liu,
\newblock ``Multiple linear regression for high efficiency video intra
  coding,''
\newblock in {\em ICASSP}, 2019.

\bibitem{harmonizing_maximum_likelihood}
Soochan Lee, Junsoo Ha, and Gunhee Kim,
\newblock ``Harmonizing maximum likelihood with {GAN}s for multimodal
  conditional generation,''
\newblock in {\em ICLR}, 2019.

\bibitem{context_aware_semantic}
Haofeng Li, Guanbin Li, Liang Lin, and Yizhou Yu,
\newblock ``Context-aware semantic inpainting,''
\newblock {\em IEEE Transactions on Cybernetics}, vol. 49, no. 12, pp.
  4398--4411, October 2019.

\bibitem{generative_adversarial_nets}
Ian Goodfellow, Jean Pouget-Abadie, Mehdi Mirza, Bing Xu, David Warde-Farley,
  Sheryl Ozair, Aaron Courville, and Yoshua Bengio,
\newblock ``Generative adversarial nets,''
\newblock in {\em NIPS}, 2014.

\bibitem{generative_image_inpainting}
Jiahui Yu, Zhe Lin, Jimei Yang, Xiaohui Shen, Xin Lu, and Thomas S.~Huang,
\newblock ``Generative image inpainting with contextual attention,''
\newblock in {\em CVPR}, 2018.

\bibitem{globally_and_locally}
Satoshi Iizuka, Edgar Simo-Serra, and Hiroshi Ishikawa,
\newblock ``Globally and locally consistent image completion,''
\newblock {\em ACM Transactions on Graphics}, vol. 36, no. 4, pp. 1--14, July
  2017.

\bibitem{pepsi_plus_plus}
Yong-Goo Shin, Min-Cheol Sagong, Yoon-Jae Yeo, Seung-Wook Kim, and Sung-Jea Ko,
\newblock ``{PEPSI}++: fast and lightweight network for image inpainting,''
\newblock {\em arXiv:1905.09010v3}, October 2019.

\bibitem{context_encoders_feature}
Deepak Pathak, Philipp Kr{\"a}henb{\"u}hl, Jeff Donahue, Trevor Darell, and
  Alexei~A. Efros,
\newblock ``Context encoders: feature learning by inpainting,''
\newblock in {\em CVPR}, 2016.

\bibitem{high_resolution_image}
Tingzhu Sun, Weidong Fang, Wei Chen, Yanxin Yao, Fangming Bi, and Baolei Wu,
\newblock ``High-resolution image inpainting based on multi-scale neural
  network,''
\newblock {\em Electronics}, vol. 8, no. 11, November 2019.

\bibitem{fully_connected_network}
Jiahao Li, Bin Li, Jizheng Xu, Ruiqin Xiong, and Wen Gao,
\newblock ``Fully-connected network-based intra prediction for image coding,''
\newblock {\em IEEE Transactions on Image Processing}, vol. 27, no. 7, pp.
  3236--3247, July 2018.

\bibitem{overview_of_the}
Gary~J. Sullivan, Jens-Rainer Ohm, Woo-Jin Han, and Thomas Wiegand,
\newblock ``Overview of the {H}igh {E}fficiency {V}ideo {C}oding ({HEVC})
  {S}tandard,''
\newblock {\em IEEE Transactions on Circuits and Systems for Video Technology},
  vol. 22, no. 12, pp. 1649--1667, December 2012.

\bibitem{intra_coding_of}
Jani Lainema, Frank Bossen, Woo-Jin Han, Junghye Min, and Kemal Ugur,
\newblock ``Intra coding of the {HEVC} standard,''
\newblock {\em IEEE Transactions on Circuits and Systems for Video Technology},
  vol. 22, no. 12, pp. 1792--1801, December 2012.

\bibitem{a_fast_hevc}
Xin Lu, Chang Yu, and Xuesong Jin,
\newblock ``A fast {HEVC} intra-coding algorithm based on texture homogeneity
  and spatio-temporal correlation,''
\newblock {\em Journal on Advances in Signal Processing}, vol. 2018, no. 37,
  June 2018.

\bibitem{context_adaptive_neural}
Thierry Dumas, Aline Roumy, and Christine Guillemot,
\newblock ``Context-adaptive neural network-based prediction for image
  compression,''
\newblock {\em IEEE Transactions on Image Processing}, vol. 29, pp. 679--693,
  August 2019.

\bibitem{progressive_spatial_recurrent}
Yueyu Hu, Wenhan Yang, Mading Li, and Jiaying Liu,
\newblock ``Progressive spatial recurrent neural network for intra
  prediction,''
\newblock {\em arXiv preprint arXiv:1807.02232}, May 2019.

\bibitem{multiple_reference_line}
B.~Bross, P.~Keydel, H.~Schwarz, D.~Marpe, T.~Wiegand, L.~Zhao, X.~Zhao, X.~Li,
  S.~Liu, Y.~Chang, H.~Jiang, P.~Lin, C.~Kuo, C.~Lin, and C.~Lin,
\newblock ``{CE}3: {M}ultiple reference line intra prediction (test 1.1.1,
  1.1.2, 1.1.3 and 1.1.4),''
\newblock {\em Joint Video Exploration Team (JVET) of ITU-T SG 16 WP 3 and
  ISO/IEC JTC 1/SC 29/WG 11, $12^{\text{th}}$ meeting, Macao}, October 2018.

\bibitem{affine_linear_weighted}
Jonathan Pfaff, Bj{\"o}rn Stallenberger, Michael Sch{\"a}fer, Philipp Merkle,
  Philipp Helle, Thobias Hinz, Heiko Schwarz, Detlev Marpe, and Thomas Wiegand,
\newblock ``Affine linear weighted intra prediction,''
\newblock {\em Joint Video Exploration Team (JVET) of ITU-T SG 16 WP 3 and
  ISO/IEC JTC 1/SC 29/WG 11, $14^{\text{th}}$ meeting, Geneva}, March 2019.

\bibitem{simplifications_of_mip}
J.~Pfaff, P.~Merkle, P.~Helle, H.~Schwarz, D.~Marpe, T.~Wiegand, A.~K.
  Ramasubramonian, T.~Biatek, G.~Van~der Auwera, L.~Pham~Van, M.~Karczewicz,
  J.~Choi, J.~Heo, J.~Lim, M.~Salehifar, S.~Kim, K.~Kondo, M.~Ikeda, T.~Suzuki,
  Z.~Zhang, K.~Andersson, R.~Sj{\"o}berg, J.~Str{\"o}m, P.~Wennersten, and
  R.~Yu,
\newblock ``Simplifications of {MIP},''
\newblock {\em Joint Video Exploration Team (JVET) of ITU-T SG 16 WP 3 and
  ISO/IEC JTC 1/SC 29/WG 11, $15^{\text{th}}$ meeting, Gothenburg}, July 2019.

\bibitem{generative_adversarial_network}
Linwei Zhu, Sam Kwong, Yun Zhang, Shiqi Wang, and Xu~Wang,
\newblock ``Generative adversarial network-based intra prediction for video
  coding,''
\newblock {\em IEEE Transactions on Multimedia}, vol. 22, no. 1, January 2020.

\bibitem{neural_network_based_intra}
J.~Pfaff, P.~Helle, D.~Maniry, S.~Kaltenstadler, W.~Samek, H.~Schwarz,
  D.~Marpe, and T.~Wiegand,
\newblock ``Neural network based intra prediction for video coding,''
\newblock in {\em SPIE}, 2018.

\bibitem{intra_picture_prediction}
Philippe Helle, Jonathan Pfaff, Michael Sch{\"a}fer, Roman Rischke, Heiko
  Schwarz, Detlev Marpe, and Thomas Wiegand,
\newblock ``Intra picture prediction for video coding with neural networks,''
\newblock in {\em DCC}, 2019.

\bibitem{intra_frame_prediction}
Fabian Brand, J{\"u}rgen Seiler, and Andr{\'e} Kaup,
\newblock ``Intra frame prediction for video coding using a conditional
  autoencoder approach,''
\newblock in {\em PCS}, 2019.

\bibitem{analysis_of_emerging}
Julien Le~Tanou and M{\'e}d{\'e}ric Blestel,
\newblock ``Analysis of emerging video codecs: coding tools, compression
  efficiency, and complexity,''
\newblock {\em SMPTE Motion Imaging Journal}, vol. 128, no. 10, pp. 14--24,
  November 2019.

\bibitem{a_hybrid_neural}
Yue Li, Li~Li, Zhu Li, Jianchao Yang, Ning Xu, Dong Liu, and Houqiang Li,
\newblock ``A hybrid neural network for chroma intra prediction,''
\newblock in {\em ICIP}, 2018.

\bibitem{combining_intra_block}
Zhaobin Zhang, Yue Li, Li~Li, Zhu Li, and Shan Liu,
\newblock ``Combining intra block copy and neighboring samples using
  convolutional neural network for image coding,''
\newblock in {\em VCIP}, 2018.

\bibitem{high_efficiency_video_coding_hevc_algorithms}
Vivienne Sze, Gary~J. Sullivan, and Madhukar Budagavi,
\newblock {\em High Efficiency Video Coding ({HEVC}): Algorithms and
  Architectures},
\newblock Springer, July 2014.

\bibitem{hevc_the_new}
Mahsa~T. Pourazad, Colin Doutre, Maryam Azimi, and Panos Nasiopoulos,
\newblock ``{HEVC}: the new gold standard for video compression: how does
  {HEVC} compare with {H.264}/{AVC}?,''
\newblock {\em IEEE Consumer Electronics Magazine}, vol. 1, no. 3, pp. 36--46,
  July 2012.

\bibitem{homogeneity_based_fast}
Mohamed Maazouz, Noureddine Batel, Nejmeddine Bahri, and Nouri Masmoudi,
\newblock ``Homogeneity-based fast {CU} partitioning algorithm for {HEVC} intra
  coding,''
\newblock {\em Engineering Science and Technology, an International Journal},
  vol. 22, no. 3, pp. 706--714, June 2019.

\bibitem{multi_type_tree}
Xiang Li, Hsiao-Chiang Chuang, Jianle Chen, Marta Karczewicz, Li~Zhang, Xin
  Zhao, and Amir Said,
\newblock ``Multi-type-tree,''
\newblock {\em Joint Video Exploration Team (JVET) of ITU-T SG 16 WP 3 and
  ISO/IEC JTC 1/SC 29/WG 11, $4^{\text{th}}$ meeting, Chengdu}, October 2016.

\bibitem{imagenet_a_large}
Jia Deng, Wei Dong, Richard Socher, Li-Jia Li, Kai Li, and Fei-Fei Li,
\newblock ``Image{N}et: a large-scale hierarchical image database,''
\newblock in {\em CVPR}, 2009.

\bibitem{an_adaptive_lagrange}
Chengyue Ma, Karam Naser, Vincent Ricordel, Patrick Le~Callet, and Chunmei
  Qing,
\newblock ``An adaptive lagrange multiplier determination method for dynamic
  texture in {HEVC},''
\newblock in {\em ICCE-China}, 2016.

\bibitem{learned_fast_hevc}
Zhibo Chen, Jun Shi, and Weiping Li,
\newblock ``Learned fast {HEVC} intra coding,''
\newblock {\em arXiv:1907.02287v1}, July 2019.

\bibitem{modified_encoder_decision}
Xin Zhao, Xiang Li, and Shan Liu,
\newblock ``{JVET}-{N}0363: {M}odified encoder decision for transform skip,''
\newblock {\em Joint Video Exploration Team (JVET) of ITU-T SG 16 WP 3 and
  ISO/IEC JTC 1/SC 29/WG 11, $14^{\text{th}}$ meeting, Geneva}, March 2019.

\bibitem{practical_recommendations_for}
Yoshua Bengio,
\newblock ``Practical recommendations for gradient-based training of deep
  architectures,''
\newblock in {\em Neural Networks: Tricks of the Trade}. 2013, pp. 437--478,
  Springer.

\bibitem{ntire_2017_challenge}
Eirikur Agustsson and Radu Timofte,
\newblock ``{NTIRE} 2017 challenge on single image super-resolution: dataset
  and study,''
\newblock in {\em CVPR Workshops}, 2017.

\bibitem{tensorflow_a_system}
M.~Abadi, P.~Barham, J.~Chen, Z.~Chen, A.~Davis, J.~Dean, M.~Devin,
  S.~Ghemawat, G.~Irving, M.~Isard, M.~Kudlur, J.~Levenberg, R.~Monga,
  S.~Moore, D.~G. Murray, B.~Steiner, P.~Tucker, V.~Vasudevan, P.~Warden,
  M.~Wicke, Y.~Yu, and X.~Zhang,
\newblock ``Tensorflow: a system for large-scale machine learning,''
\newblock in {\em OSDI}, 2016.

\bibitem{common_test_conditions}
Franck Bossen,
\newblock ``Common test conditions and software reference configurations,''
\newblock in {\em Proceedings of the $12^{\text{th}}$ {JVT}-{VC} meeting,
  {JCTVC}-{K}1100, Shanghai, {CN}}, October 2012.

\bibitem{calculation_of_average}
Gisle Bjontegaard,
\newblock ``Calculation of average {PSNR} differences between {RD}-curves,''
\newblock Tech. {R}ep., ITU-T SG16/Q6, Austin, TX, USA, Tech. Rep. VCEG-M33,
  April 2001.

\bibitem{kodak_suite}
``Kodak suite,'' \url{http://r0k.us/graphics/kodak/},
\newblock accessed: 2019-11-04.

\bibitem{a_database_of}
David Martin, Charless Fowlkes, Doron Tal, and Jitendra Malik,
\newblock ``A database of human segmented natural images and its application to
  evaluating segmentation algorithms and measuring ecological statistic,''
\newblock in {\em ICCV}, 2001.

\bibitem{intra_sub_partitions}
Santiago De-Lux{\'a}n-Hern{\'a}ndez, Valeri George, Jackie Ma, Tung Nguyen,
  Heiko Schwarz, Detlev Marpe, and Thomas Wiegand,
\newblock ``{CE}3: Intra sub-partitions coding mode (tests 1.1.1 and 1.1.2),''
\newblock {\em Joint Video Exploration Team (JVET) of ITU-T SG 16 WP 3 and
  ISO/IEC JTC 1/SC 29/WG 11, $13^{\text{th}}$ meeting, Marrakech}, January
  2019.

\bibitem{versatile_video_coding_draft_5}
Benjamin Bross, Jianle Chen, and Shan Liu,
\newblock ``Versatile {V}ideo {C}oding ({D}raft 5),''
\newblock {\em Joint Video Exploration Team (JVET) of ITU-T SG 16 WP 3 and
  ISO/IEC JTC 1/SC 29/WG 11, $14^{\text{th}}$ meeting, Geneva}, March 2019.

\bibitem{versatile_video_coding_draft_6}
Benjamin Bross, Jianle Chen, and Shan Liu,
\newblock ``Versatile {V}ideo {C}oding ({D}raft 6),''
\newblock {\em Joint Video Exploration Team (JVET) of ITU-T SG 16 WP 3 and
  ISO/IEC JTC 1/SC 29/WG 11, $15^{\text{th}}$ meeting, Gothenburg}, July 2019.

\bibitem{quadtree_plus_binary}
Jicheng An, Han Huang, Kai Zhang, Yu-Wen Huang, and Shawmin Lei,
\newblock ``Quadtree plus binary tree structure integration with {JEM} tools,''
\newblock {\em Joint Video Exploration Team (JVET) of ITU-T SG 16 WP 3 and
  ISO/IEC JTC 1/SC 29/WG 11, $2^{\text{nd}}$ meeting, San Diego}, February
  2016.

\bibitem{highly_flexible_coding}
Fabrice Le~L{\'e}annec, Tangi Poirier, Franck Galpin, Fabrice Urban,
  {\'E}douard Fran{\c c}ois, Wei-Jung Chien, Vadim Seregin, and Marta
  Karczewicz,
\newblock ``Highly flexible coding structures for next-generation video
  compression standard,''
\newblock in {\em DCC}, 2019.

\bibitem{efficient_and_accurate}
Xiangyu Zhang, Jianhua Zou, Xiang Ming, Kaiming He, and Jian Sun,
\newblock ``Efficient and accurate approximations of nonlinear convolutional
  neural networks,''
\newblock in {\em CVPR}, 2015.

\bibitem{nips_pruning_networks}
R.~Yu, A.~Li, C.~Chen, J.~Lai, V.~I. Morariu, X.~Han, M.~Gao, C~Lin, and L.~S.
  Davis,
\newblock ``{NIPS}: pruning networks using neuron importance score
  propagation,''
\newblock in {\em CVPR}, 2018.

\end{thebibliography}

\end{document}